\documentclass[fleqn,aps,prb,twocolumn,showpacs,nofootinbib,floatfix,longbibliography]{revtex4-2}
\usepackage{amsmath,amssymb,amsfonts,float,graphics,epsfig,epstopdf,color,verbatim,tabularx,bm,multirow,appendix,hyperref}
\usepackage{amsmath}
\usepackage{amssymb}
\usepackage{mathtools}
\usepackage{graphicx}
\usepackage{lipsum}

\usepackage{bm}
\usepackage{hyperref}
\usepackage{url}
\usepackage[utf8]{inputenc}
\usepackage{subfigure}
\usepackage{slashed,bbm}
\usepackage{graphics,psfrag,epsfig}
\usepackage{dsfont}
\usepackage{setspace}
\usepackage{wasysym}
\usepackage{slashed}
\usepackage{lipsum}
\usepackage{braket}
\usepackage{physics}
\usepackage{enumerate}%

\usepackage{hyperref}
\usepackage{xcolor}

\definecolor{dark-red}{rgb}{0.4,0.15,0.15}
\definecolor{dark-blue}{rgb}{0.15,0.15,0.4}
\definecolor{medium-blue}{rgb}{0,0,0.5}
\hypersetup{
colorlinks, linkcolor={dark-blue},
citecolor={dark-blue}, urlcolor={medium-blue}
}
\usepackage{ulem}
\newcommand{\be}{\begin{equation}}
\newcommand{\ee}{\end{equation}}
\newcommand{\bea}{\begin{eqnarray}}
\newcommand{\eea}{\end{eqnarray}}

\renewcommand{\i}{\text{i}}

\begin{document}

\title{Unconventional topological mixed-state transition and critical phase induced by self-dual coherent errors}

\author{Yu-Hsueh Chen}
\affiliation{Department of Physics, University of California at San Diego, La Jolla, California 92093, USA}
\author{Tarun Grover}
\affiliation{Department of Physics, University of California at San Diego, La Jolla, California 92093, USA}

\begin{abstract}
A topological phase can undergo a phase transition driven by anyon condensation. A potential obstruction to such a mechanism could arise if there exists a symmetry between anyons that have non-trivial mutual statistics. Here we consider toric code subjected to errors that tend to proliferate anyons with non-trivial mutual statistics. Using triangle inequality, we show that in the presence of electromagnetic duality and a partial-transpose symmetry, a decoherence induced phase transition out of the topological phase must be rather unconventional and lie beyond standard rules of anyon condensation. To explore such physics, we first subject toric code to a self-dual quantum channel where Kraus operators are proportional to $X+Z$. We find that the topological phase is stable up to the maximal error rate, when viewing density matrix as a pure state in the double Hilbert space.  To access an unconventional transition, we then consider a perturbed toric code subjected to the  self-dual channel, and find numerical evidence that beyond a critical error rate, the topological phase is destroyed resulting in a critical phase where anyons are only power-law condensed.
\noindent

\end{abstract}

\maketitle


\section{Introduction}
\noindent

Topological phases of matter host anyons that can have non-trivial mutual statistics with respect to each other \cite{wen2004quantum}. The anyonic statistics is not only responsible for the universal properties of a gapped, topological phase, it also puts strong constraints on the nature of quantum criticality and the proximate phases. For example, condensing an anyon confines all other anyons that have non-trivial mutual statistics with respect to it \cite{burnell2018anyon, bais2002broken, bais2007breaking, bais2009condensate,neupert2016boson}. Relatedly, one cannot condense two anyons simultaneously that have non-trivial mutual statistics. This can lead to phases of matter or critical points that are difficult to  understand in terms of anyon condensation \cite{jongeward1980monte,tupitsyn2010topological,vidal2008low,wu2012phase,zhu2019gapless, somoza2021self}. Motivated by such considerations, in this paper we study \textit{mixed states} obtained by subjecting topologically ordered pure states to quantum channels that have a tendency to condense anyons with non-trivial mutual statistics. For a 2+1-D $\mathbb{Z}_2$ topologically ordered state, we derive certain general constraints that imply the decoherence driven phase transition that destroys the topological order must violate the standard rules of anyon condensation in one way or another. Specifically for the toric code, we find numerical evidence for a phase transition out of the topological state to a critical phase of matter in which anyons are `algebraically condensed' i.e. the associated string operators decay as a power law.

\begin{figure}
	\centering
	\includegraphics[width=0.9\linewidth]{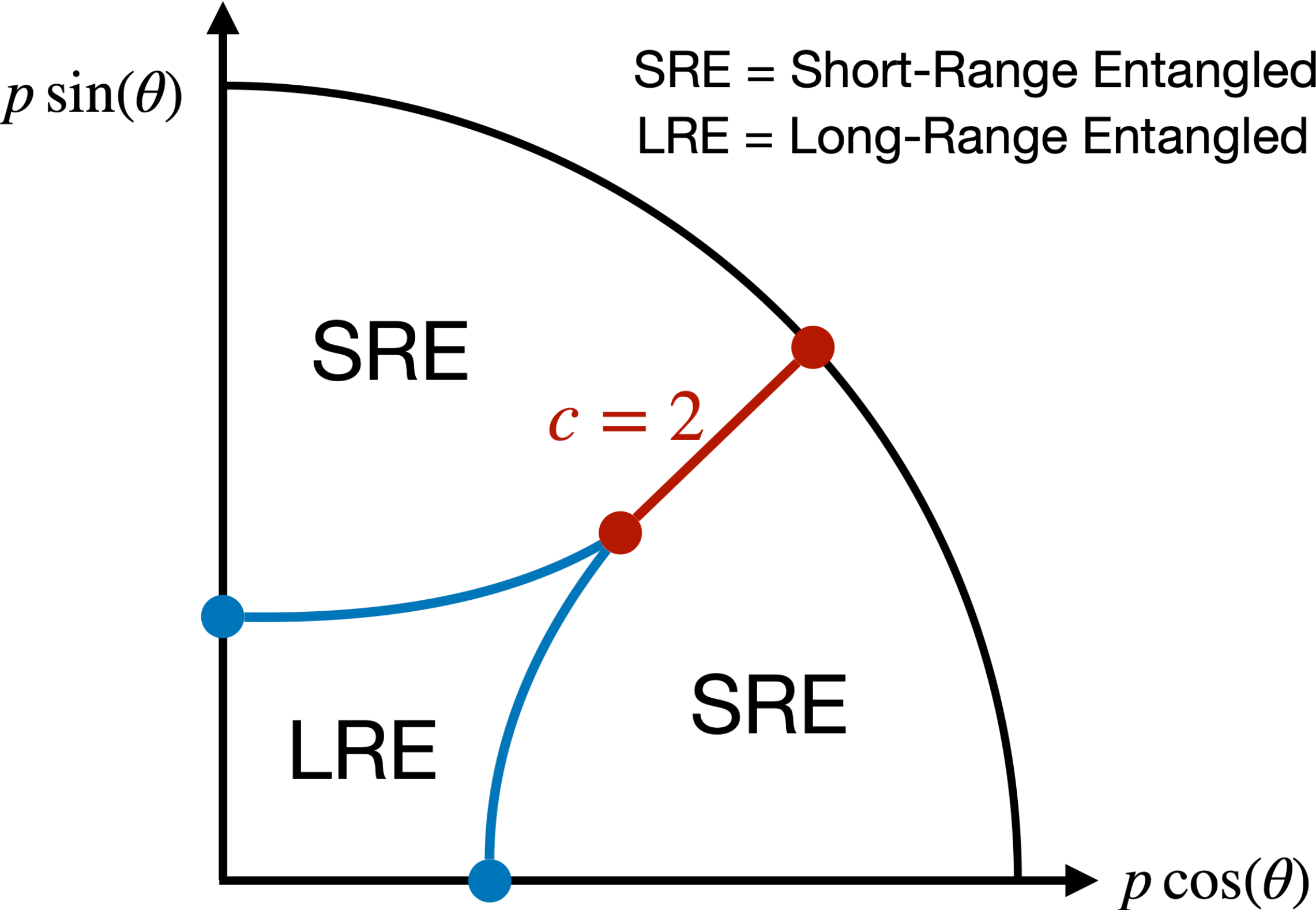}
	\caption{ Schematic phase diagram of the mixed state studied in this work based on the 2nd Renyi entropy. The mixed state is obtained by subjecting the pure state $|\Psi (h) \rangle = \prod_e [I + h(X_e + Z_e)/\sqrt{2}]|\text{Toric code}\rangle$ to the coherent channel $\mathcal{E}_e[\cdot] = (1-p)(\cdot) + p \sigma_e(\theta)(\cdot) \sigma_e (\theta)$, where $\sigma_e (\theta) = \cos(\theta)Z_e + \sin(\theta) X_e$. We focus primarily on the self-dual line (i.e. $\theta = \pi/4$). The entanglement along the red line is consistent with a central charge $c = 2$ conformal field theory. The plot shown here corresponds to $h = 0.2$.
	}
	\label{Fig:fig1}
\end{figure}

To motivate our setup, let us recall the phase diagram of the toric code Hamiltonian \cite{kitaev2003fault}, $H_{TC}$, perturbed by uniform fields along $X$ and $Z$ directions in the spin-space: $H = H_{\text{TC}} - h_x \sum_i Z_i - h_z \sum_i X_i $ \cite{wegner1971duality,fradkin1979phase,jongeward1980monte,tupitsyn2010topological,vidal2008low,wu2012phase,somoza2021self}. When $h_x > h_z$ ($h_x < h_z$), increasing $h_x$ ($h_z$) leads to a second-order phase transition that can be understood as a condensation of a self-boson anyon, namely, either the electric charge $e$ or the magnetic flux $m$ \cite{fradkin1979phase}. However, along the self-dual line $h_x = h_z$, the mass gap to the $e$ and the $m$ anyon is equal, and due to their $\pi$ mutual statistics, the fate of the ground state is not so obvious as one leaves the topological phase. Numerically, one finds an intriguing multicritical point along the $h_x = h_z$ line as one exits the topological phase \cite{jongeward1980monte,tupitsyn2010topological,vidal2008low,wu2012phase,somoza2021self} whose universal exponents are approximately known, but the corresponding field theory is not well-understood \cite{somoza2021self}. 

Let us now consider a different setup where the toric code ground state is subjected to \textit{local decoherence}, so that the system is described by a mixed state. For example, let’s first subject the ground state to a channel with local Kraus operators $\propto X$ whose action on the ground state generates a pair of electric anyons $e$ (in the convention where the plaquette term in the toric code Hamiltonian is  $\prod_{e \in p} X_e$ while the star term is $\prod_{e \in v} Z_e$). Increasing the strength of the decoherence leads to a phase transition to a non-topological phase which can be characterized in multiple ways \cite{dennis2002,wang2003confinement, lee2023quantum, fan2023diagnostics,bao2023mixed,chen2023separability,sang2023mixed,li2024replica,su2024tapestry, lee2024exact,lyons2024understanding}. One way to understand this transition is via the `double-state formalism' \cite{ lee2022symmetry,lee2023quantum,bao2023mixed}, in which the density matrix is mapped to a pure state $|\rho\rangle$ that lives in the tensor product of the original Hilbert space with its own copy \cite{schmutz1978real,prosen2008third,jamiolkowski1972linear, choi1975completely}. One can show that quantities linear in density matrix cannot detect local decoherence induced phase transitions, and one of the simplest quantities that does capture such transitions is the normalization of the double state $\langle \rho|\rho\rangle$, which equals the second Renyi entropy of the mixed state \cite{lee2023quantum,fan2023diagnostics,bao2023mixed}. Furthermore, if the double state $|\rho\rangle$ is short-ranged-entangled (SRE), then the density matrix $\propto \rho^2 \otimes I$ can be expressed as a convex sum of SRE states \cite{chen2023symmetry}. Although the universality class or the threshold for the transition within the double state formalism is not identical to that for the intrinsic mixed-state transition for a single copy of the system \cite{dennis2002, lee2023quantum, fan2023diagnostics,chen2023separability,sang2023mixed,lee2024exact}, it still provides an intuitive understanding of the transitionIn particular, the phase transition for the double state corresponds to condensing the bound state $e \bar{e}$ where $e$ and $\bar{e}$ refer to the (electric) charge anyons in the two copies respectively \cite{lee2023quantum,bao2023mixed}. Similarly, subjecting the toric code ground state to local Kraus operators $\propto Z$ also destroys the double topologically ordered  state and the corresponding transition can now be understood as condensing the bound state $m \bar{m}$ of anyons where $m$ and $\bar{m}$ refer to the (magnetic) flux anyons in the two copies respectively \cite{lee2023quantum,bao2023mixed}. If one now simultaneously subjects the system to the composition of the aforementioned two channels, the resulting transition can also be understood in terms of anyon condensation, since $e \bar{e}$ and $m \bar{m}$ are mutual bosons. What happens if one instead considers  a quantum channel with Kraus operators of the form $X+Z$? Now it is not obvious that one can condense either $e \bar{e}$ or $m \bar{m}$ anyons, since the probability amplitude for condensing $e$ or $m$ in either copy is equal due to the symmetry of the channel. This is reminiscent to the aforementioned (pure) toric code ground state along the self-dual line. Such Kraus operators will be the focus of our study. 

To illustrate the special properties of the Kraus operator $X+Z$, we first ask a general question: what conditions must the Kraus operators satisfy such that a conventional phase transition out of the topologically ordered double state of the toric code is not even allowed? Here by `conventional' we mean a transition that corresponds to condensation of a self-boson (such as $e\bar{e}$) and where the standard rules of anyon condensation apply, e.g., one expects that the string operator corresponding to the condensing anyon has long-range order, while anyons that have non-trivial mutual statistics with the condensed anyon confine. In Sec.\ref{sec:condition} we show that if a channel possesses electromagnetic duality (EMD) symmetry and a `partial-transpose symmetry' (which essentially corresponds to sending $X \rightarrow X, Z \rightarrow Z, Y \rightarrow -Y$ in only one copy of the double state), then any transition out of the topological phase must be unconventional. We show this result using Cauchy-Schwarz (i.e. triangle) inequality. The basic idea is to express anyon condensation rules in terms of asymptotic behavior of certain matrix elements between excited states that correspond to the action of anyon-creating string operators on the ground state. Using EMD and partial-transpose symmetry,  matrix elements corresponding to physically distinct processes can be related  to each other. At the same time, one can use  triangle inequality to bound the product of matrix elements in terms of other matrix elements. If one now assumes that the phase transition out of the double topologically ordered  state happens due to condensation of a self-boson, one can then show that EMD symmetry along with the partial-transpose symmetry leads to a contradiction with the triangle inequality.  On that note, the density matrix obtained by decoherence with local Kraus operators $X + Z$ possesses both EMD and partial-transpose symmetry, which motivate us to study this specific channel in detail.

To illustrate the implications of the aforementioned constraints on topological phase transitions due to EMD and partial-transpose symmetry, we start with a class of pure states of the form $
|\Psi(h)\rangle =  \prod_e (I + h (X_e + Z_e)/\sqrt{2})   |\textrm{Toric code ground state}\rangle$ where $h \in [0,1]$. Such a pure state was first studied in Ref.\cite{zhu2019gapless} where an interesting critical phase for a range of $h$ was identified. We subject these class of pure states to Kraus operators of the form $X \cos(\theta)+Z \sin(\theta)$. When $\theta = \pi/4$, the density matrix respects both EMD and partial-transpose symmetry. We then use the aforementioned double state formalism to map the problem to a non-unitarily evolved pure state, where the evolution time depends on the strength of the decoherence. The normalization of the double state can be interpreted as a partition function of a statistical mechanics problem whose phase diagram reveals different phases of the mixed state, analogously to the case of $Z$ or $X$ decoherence  \cite{lee2023quantum,bao2023mixed}.  By employing this mapping, we analyze the phase diagram for our problem with a combination of numerical techniques and analytical arguments. The result is schematically shown in Fig.\ref{Fig:fig1}. We find that along the self-dual line, i.e., $\theta = \pi/4$, the double state undergoes a phase transition from a topological ordered phase to a phase with \textit{critical} (i.e. power-law) correlations. The extent of the critical phase depends on the parameter $h$ in the non-decohered original state $|\Psi(h)\rangle $. When $h = 0$, i.e., when the original state is simply the toric code, the correlation functions along the self-dual line can be exactly mapped to the self-dual Ashkin-Teller (AT) model, and it turns out that in this special case the critical phase shrinks to a single point whose properties correspond to a single, gapless complex scalar in $1+1$-D i.e., central charge $c = 1$ theory. In contrast, when $h > 0$, one finds a \textit{critical phase} for a whole range of decoherence strength whose entanglement is consistent with a central charge $c = 2$ theory. We also find a relation of the corresponding statistical mechanics model with two copies of the AT model along the self-dual line which suggests  that the low-energy theory of this phase is simply two copies of a gapless, complex scalar.

The remaining part of the paper is organized as follows.
In Sec.\ref{sec:condition}, we discuss the conditions under which a channel guarantees that a mixed-state transition (if it exists) can not be described by a standard anyon condensation scheme. 
In Sec.\ref{sec:coherent}, we consider a self-dual coherent channel that  satisfies the conditions in Sec.\ref{sec:condition} and study its effect on the toric code ground state.
In Sec.\ref{sec:nonfixed_coherent}, we  apply the self-dual channel to a non-fixed point toric code and obtain the phase diagram schematically sketched in Fig.\ref{Fig:fig1}. We conclude in Sec.\ref{sec:discuss} with a summary and discussion of our results.

\section{Constraints on anyon condensation from self-duality and partial-transpose symmetry}
\label{sec:condition}

The standard anyon condensation scheme leads to various constraints for possible gapped ground states proximate to a given topological phase \cite{burnell2018anyon, bais2002broken, bais2007breaking, bais2009condensate,neupert2016boson}. For example, condensing a self-boson anyon confines all particles that have non-trivial mutual statistics with it. In this section we will study constraints from the EMD symmetry and a partial-transpose symmetry (introduced below) on anyon condensation in a decohered toric code, which we will serve as a guide to search for decoherence induced novel transitions that must lie beyond the standard anyon condensation scheme.

\begin{figure}
	\centering
	\includegraphics[width=\linewidth]{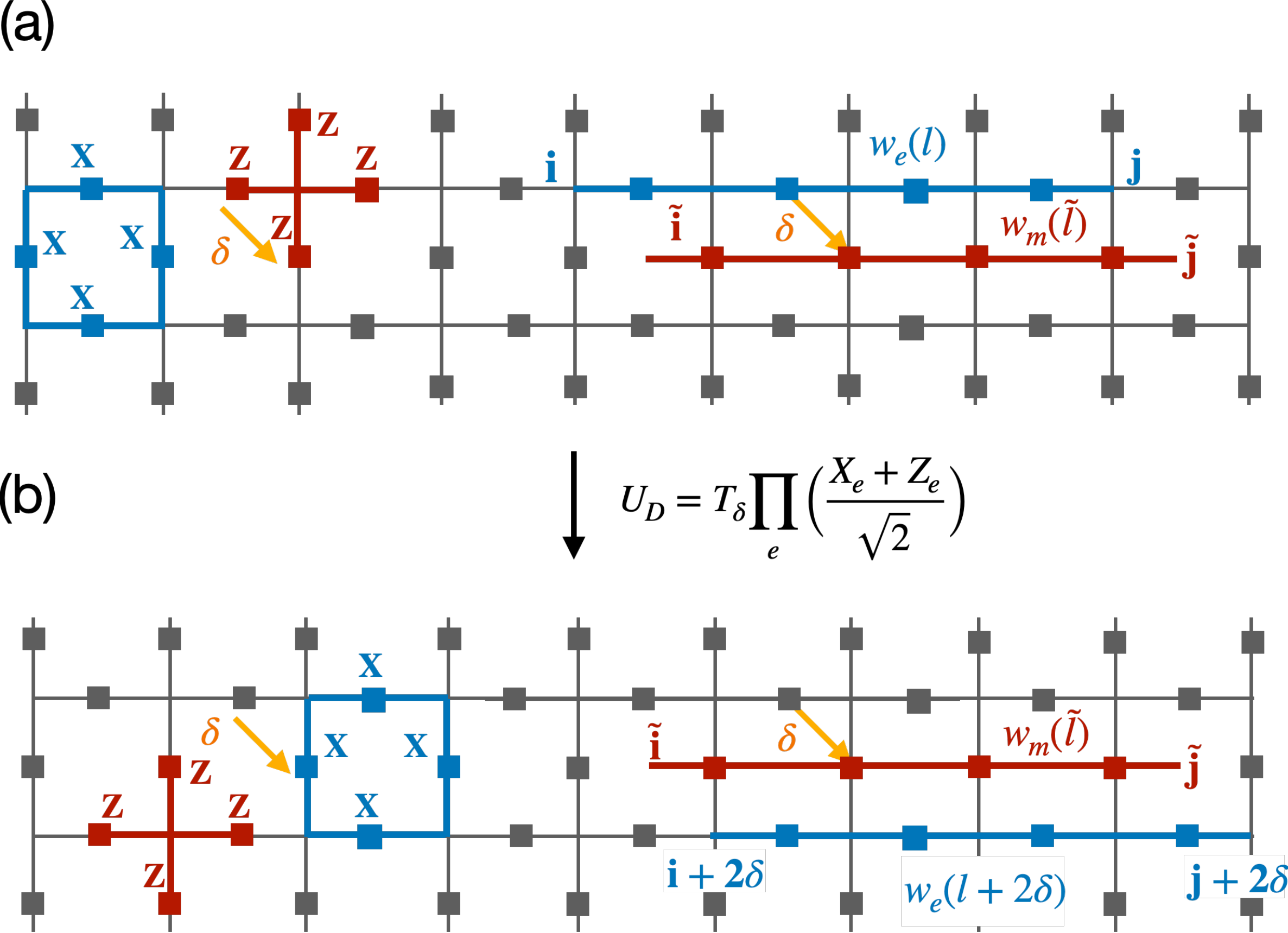}
	\caption{ (a) Hamiltonian of the toric code and the string operator $w_{e}(l) [w_m(\tilde{l})]$ that creates a pair of charge (flux) anyons at sites $i (\tilde{i})$ and $j (\tilde{j})$.
		The orange arrows depict the translation along the diagonal direction.
		(b) The resulting Hamiltonian and the string operators after conjugating with the EMD unitary $U_D = T_\delta \prod_e [(X_e + Z_e)/\sqrt{2}]$.
	} 
	\label{Fig:toric_model}
\end{figure}

Let us first introduce our notation for anyon-creating string operators that we will employ extensively.
The Hamiltonian of the toric code can be written as 
\begin{equation}
	H_\text{TC} = - \sum_v A_v- \sum_p B_p,
\end{equation}
where $A_v = \prod_{e \in v} Z_e$, $B_p = \prod_{e \in p} X_e$, and
the qubits reside on the edges (denoted as $e$) of a square lattice (Fig.\ref{Fig:toric_model}).
The ground state $ |\Psi_0\rangle$ satisfies $A_v |\Psi_0\rangle = B_p |\Psi_0\rangle = |\Psi_0\rangle,\ \forall v,p$. 
A pair of charge anyons $|e_i e_j\rangle$ located at site $i$ and $j$ correspond to the violation of $A_v$ on sites $i$ and $j$, and can be created by applying $w_e(l) = \prod_{e \in l} X_e$, a Pauli-$X$ string $l$ with the end points located at sites $i$ and $j$, to $|\Psi_0\rangle$ (see Fig.\ref{Fig:toric_model}).
Similarly, a pair of flux anyons $|m_{\tilde{i}} m_{\tilde{j}}\rangle$ located on the dual lattice with site $\tilde{i}$ and $\tilde{j}$ corresponds to the violation of $B_p$ on sites $\tilde{i}$ and $\tilde{j}$, and can be created by applying $w_m(\tilde{l}) = \prod_{e \in \tilde{l}} Z_e$, a Pauli-$Z$ string $\tilde{l}$ with the end points located at sites $\tilde{i}$ and $\tilde{j}$, to $|\Psi_0\rangle$.

The toric code Hamiltonian possesses an electro-magnetic duality (EMD) symmetry corresponding to the unitary transformation
\begin{equation}
	U_D = T_\delta \prod_e \Big( \frac{X_e + Z_e}{\sqrt{2}}\Big) =  T_\delta \prod_e \sigma_e, \label{eq:U_D}
\end{equation}
where $T_\delta$ is a translation operator along the diagonal direction [{see Fig.\ref{Fig:toric_model}}(a)].
The involvement of the translation operator in $U_D$ leads to some subtleties which we now discuss.
First, conjugating  $U_D$ with the aforementioned charge and flux creation operators yields
\begin{equation}
	U_D w_e(l) U_D^\dagger = w_m(\tilde{l}),\ U_D w_m(\tilde{l}) U_D^\dagger = w_e({l} + 2 \delta)
\end{equation}
[see Fig.\ref{Fig:toric_model}(b)].
Therefore, EMD transformation exchanges the charge and flux operators up to a lattice translation.
One may then naively think that the fermion creation operator $w_f(l) \equiv w_e(l) w_m(\tilde{l})$ is invariant under the EMD transformation up to a lattice translation (we define fermions on the original lattice without loss of generality).
However, a straightforward algebra shows that 
\begin{equation}
	\label{Eq:fermion_conjugate}
	\begin{aligned}
		U_D w_f(l)  U_D^\dagger & = (U_D w_e(l) U_D^\dagger )( U_D w_m(\tilde{l})  U_D^\dagger )  \\
		&= w_e(l+2 \delta) w_m(\tilde{l}),
	\end{aligned}
\end{equation}
which is not equal to 
$w_f(l+2\delta) = w_e(l+2 \delta) w_m (\tilde{l} + 2 \delta)$.
Indeed, conjugating $w_f(l)$ with $U_D$ not only shifts the locations of the fermions by $2 \delta$ but also rotates them by an  angle $\pi$.
Second, since we will be interested in applying  a local quantum channel $\mathcal{E}$ to the toric code,  one can ask whether $\mathcal{E}$ respects the ``strong" or ``weak" symmetry of $U_D$ \cite{de2022symmetry}.
Following the definition in Ref.\cite{de2022symmetry}, 
a channel $\mathcal{E}$  respects the strong symmetry of $U_D$ if  each of its Kraus operator individually commutes with $U_D$ up to a
constant phase factor while only respects the weak symmetry if $ U^\dagger_D\mathcal{E}[U_D (\cdot)U_D^\dagger]U_D = \mathcal{E}[\cdot]$ .
Since $U_D$ involves translation and we require $\mathcal{E}$ to be the composition of local channels (we assume that the decoherence is generated by local errors), $\mathcal{E}$ can only respect the weak symmetry of $U_D$.

Based on the intuition that one can't simultaneously condense anyons with non-trivial mutual statistics, one might naively think that a weak EMD symmetry is already enough to rule out standard anyon condensation transition for a toric code subjected to local decoherence. This is not true however. A channel that satisfies the weak EMD symmetry is the composition of bit-flip and phase-flip channels on all edges with equal error rate. It has already been shown in Ref.\cite{fan2023diagnostics} that for toric code, the bit-flip and phase-flip channels act independently on the density matrix, and the statistical mechanics model for the transition is simply two decoupled random-bond Ising models along the Nishimori line.
Therefore, the weak EMD symmetry is not enough to induce novel mixed-state transitions.   
In fact, this can be easily understood by mapping the density matrix into a state vector in the double Hilbert space \cite{lee2023quantum,bao2023mixed}. Denoting the charge (flux) anyons in the two copies as $e (m)$ and $\bar{e} (\bar{m})$, the effect of the simultaneous bit-flip and phase-flip errors then corresponds to condensing mutual bosons, $m \bar{m}$ and $e \bar{e}$ simultaneously \cite{bao2023mixed,lee2023quantum}, which does not violate any anyon condensation rule.
Based on this intuition, it is reasonable to search for  additional conditions that may lead to stronger constraints that prohibit standard anyon condensation transition in the double Hilbert space. 

To keep the discussion self-contained, we will first briefly review the double-state formalism introduced in Ref.\cite{lee2022symmetry,lee2023quantum,bao2023mixed}, along with 
an overview of order parameters to detect anyon condensation and confinement \cite{schuch2013topological,haegeman2015shadows, duivenvoorden2017entanglement}. By employing this notation, we will soon argue that if one imposes a `partial-transpose symmetry' defined below in addition to the EMD symmetry, then the double state corresponding to the decohered toric code is not allowed to undergo a conventional anyon condensation transition.

Given a density matrix $\rho_\mathcal{H}$ acting on the Hilbert space $\mathcal{H}$, one can define a state vector (which we refer to as ``double state") in the doubled Hilbert space $\mathcal{H} \otimes \bar{\mathcal{H}}$ using Choi-Jamiolkowski (C-J) map \cite{schmutz1978real,prosen2008third,jamiolkowski1972linear, choi1975completely}: $	|\rho \rangle_{\mathcal{H} \otimes \bar{\mathcal{H}}} \equiv \rho_\mathcal{H} \otimes I_{\bar{\mathcal{H}}} | \Phi\rangle_{\mathcal{H} \otimes \bar{\mathcal{H}}}$,
where $ | \Phi\rangle_{\mathcal{H} \otimes \bar{\mathcal{H}}}$ is a product of maximally entangled pairs connecting $\mathcal{H}$ and $\bar{\mathcal{H}}$.
Henceforth, for notational simplicity, we omit the subscripts labeling the Hilbert space.
After applying the C-J map, one finds that  the initial pure state density matrix $\rho_0 = |\Psi_0\rangle \langle \Psi_0 |$ is mapped to
$|\rho_0\rangle  = |\Psi_0\rangle |\bar{\Psi}_0\rangle$ while the channel $\mathcal{E}[\cdot] = \sum_n K_n (\cdot) K_n^\dagger$ is mapped to the operator $\mathcal{E} = \sum_n K_n \otimes \bar{K}_n$.
Therefore, the non-decohered (initial) state $|\rho_0\rangle$ has double topological order (e.g. the topological entanglement entropy will be $2 \log(2)$), and the decohered density matrix $\rho = \mathcal{E}[\rho] $ after the CJ map corresponds to evolving $|\rho_0\rangle$ through the non-unitary operator $\mathcal{E}$, i.e., $|\rho\rangle = \mathcal{E}|\rho_0\rangle$.
It is also natural to consider the error-corrupted anyon states
$	|\rho_{\alpha \bar{\alpha}}\rangle =\mathcal{E} |\rho_{\alpha \bar{\alpha},0}\rangle= \mathcal{E} w_\alpha (l) \bar{w}_\alpha(l) |\rho_0\rangle $,
where $w_\alpha(l)$ is a string operator along the path $l$ that creates a pair of anyons at the end points of $l$.
To detect topological transitions in the double Hilbert space, one can consider the following $\alpha \bar{\alpha}$ anyon condensation parameter \cite{bao2023mixed}:
\begin{equation}
	\label{Eq:double_condense}
	\langle I \bar{I} | \alpha \bar{\alpha}\rangle \equiv
	\lim_{| i - j| \rightarrow \infty} \frac{\langle \rho |\rho_{\alpha \bar{\alpha}} \rangle}{\langle \rho |\rho\rangle}
	= \lim_{| i - j| \rightarrow \infty} \frac{\langle \rho_0 |\mathcal{E}^\dagger \mathcal{E} |\rho_{\alpha \bar{\alpha},0} \rangle}{\langle \rho_0 | \mathcal{E}^\dagger \mathcal{E}|\rho_0\rangle},
\end{equation}
where $i$ and $j$ denote the end points of the string $w_\alpha (l)$.
Physically, $\langle I \bar{I} | \alpha \bar{\alpha}\rangle$ detects whether the states $|\rho_0\rangle$ and $|\rho_{\alpha \bar{\alpha}, 0} \rangle$ are still physically distinct after evolving both of them through the non-unitary operator $\mathcal{E}$. 
Therefore, $\langle I \bar{I} | \alpha \bar{\alpha}\rangle $  saturates to a finite constant when $\alpha \bar{\alpha}$ is condensed. This will lead to the confinement of any anyons $\beta \bar{\gamma}$ that have non-trivial mutual statistics with  $\alpha \bar{\alpha}$ \cite{bais2009condensate,burnell2018anyon}, which can be detected through the following anyon confinement parameter  \cite{schuch2013topological,haegeman2015shadows, duivenvoorden2017entanglement}:
\begin{equation}
	\label{Eq:double_confine}
	\langle \beta \bar{\gamma}| \beta \bar{\gamma} \rangle \equiv \lim_{| i - j| \rightarrow \infty} \frac{\langle \rho_{ \beta \bar{\gamma}} |\rho_{ \beta \bar{\gamma}} \rangle}{\langle \rho |\rho\rangle},
\end{equation}
where $|\rho_{\beta \bar{\gamma}} \rangle = \mathcal{E}|\rho_{\beta \bar{\gamma},0} \rangle = \mathcal{E} w_\beta(l) \bar{w}_\gamma(l) |\rho_0\rangle=  \mathcal{E}  |\Psi_\beta, \bar{\Psi}_{\gamma}\rangle$. Intuitively, $\langle \rho_{ \beta \bar{\gamma}} |\rho_{ \beta \bar{\gamma}} \rangle$ for a long string $l$ detects the probability for processes that lead to a well-separated pair of the bound state $\beta \bar{\gamma}$. Confinement of $\beta \bar{\gamma}$ implies that this amplitude decays exponentially with $|l| = |i-j|$, and thus $\langle \beta \bar{\gamma}| \beta \bar{\gamma} \rangle = 0$ when the anyon $\beta \bar{\gamma} $ is confined. 
From Eq.\eqref{Eq:double_condense} and Eq.\eqref{Eq:double_confine}, one observes that the most general quantities of interest are $\langle {\alpha \bar{\beta}} | {\gamma \bar{\delta}} \rangle \equiv \langle \rho_{\alpha \bar{\beta}} | \rho_{\gamma \bar{\delta}} \rangle / \langle \rho | \rho\rangle$, the ``overlap" between the error-corrputed anyon eigenstates.
In the original Hilbert space, these quantities are all related to the second moment of the density matrix.
We also note that by definition, the order parameters $\langle {\alpha \bar{\beta}} | {\gamma \bar{\delta}} \rangle$ are \textit{not} conventional physical observables of $|\rho\rangle$, i.e., they cannot be written as $\langle \rho | O|\rho\rangle/\langle \rho|\rho\rangle$ where $O$ is a \textit{state-independent} operator. However, one may formulate $\langle {\alpha \bar{\beta}} | {\gamma \bar{\delta}} \rangle$ as the expectation value of a state-dependent operator $O^{\alpha \bar{\beta}}_{\gamma \bar{\delta}} \equiv ( \mathcal{E}^\dagger)^{-1} w_\alpha^\dagger w_{\bar{\beta}}^\dagger \mathcal{E}^\dagger \mathcal{E}w_\gamma w_{\bar{\delta}} \mathcal{E}^{-1} + h.c.$ \textit{as long as} the inverse of $\mathcal{E}$ exists.
This implies that these order parameters may not be a faithful measure of the underlying phase of matter when $\mathcal{E}$ has non-trivial kernel.
We will later encounter such a situation in Sec.\ref{sec:coherent}.

We now propose the conditions under which a decoherence induced transition in toric code cannot proceed via condensing a self-boson, and which therefore guarantees a mixed-state transition (if any) beyond the standard anyon condensation scheme.
Consider the decohered double state $|\rho\rangle = \mathcal{E}|\rho_0\rangle =\mathcal{E}|\Psi_0 , \bar{\Psi}_0\rangle $ with $|\Psi_0\rangle$ a $\mathbb{Z}_2$ topologically ordered state respecting EMD and time-reversal (i.e., complex conjugation) symmetry.
If the C-J transformed channel $\mathcal{E}$ satisfies the following properties, then any transition of $|\rho\rangle$ out of the double $\mathbb{Z}_2$ topologically ordered state can't be described via standard anyon condensation scheme.

\begin{enumerate}[(i)]
	\item  $\mathcal{E}$ respects the EMD symmetry:  ${\mathbb{U}_D} \mathcal{E} {\mathbb{U}_D^\dagger} = \mathcal{E}$ where $\mathbb{U}_D \equiv U_\text{D} \bar{U}_\text{D}$.
	\item $\mathcal{E}^\dagger \mathcal{E}$  is invariant under partial transpose on either $\mathcal{H}$ and $\bar{\mathcal{H}}$, i.e., $({\mathcal{E}}^\dagger\mathcal{E})^{T_\mathcal{H}} = ({\mathcal{E}}^\dagger\mathcal{E})^{\bar{T}_{\bar{\mathcal{H}}} } = {\mathcal{E}}^\dagger\mathcal{E}$,
	
\end{enumerate}

\noindent where we denote $T_\mathcal{H}$ and $\bar{T}_{\bar{\mathcal{H}}}$ as the partial transpose in the Hilbert space $\mathcal{H}$ and $\bar{\mathcal{H}}$, respectively. We note that the condition (i) is nothing but the C-J transformed version of the aforementioned weak EMD symmetry for $\mathcal{E}$. To see the implication of condition (ii), let us introduce the notation $|\Psi_\alpha\rangle \equiv w_\alpha(l)|\Psi_0\rangle$ for the $\alpha$ anyon state. Since $|\Psi_{\alpha}\rangle$ and $|\bar{\Psi}_\beta \rangle$ are time-reversal symmetric, condition (ii) implies 
\begin{equation}
	\label{Eq:partial_TR}
	\begin{aligned}
		\langle \Psi_\alpha, \bar{\Psi}_\beta|\mathcal{E}^\dagger \mathcal{E}|\Psi_\gamma, \bar{\Psi}_\delta\rangle & = \langle \Psi_\gamma, \bar{\Psi}_\beta|\mathcal{E}^\dagger \mathcal{E}|\Psi_\alpha, \bar{\Psi}_\delta\rangle \\
		& = \langle \Psi_\alpha, \bar{\Psi}_\delta |\mathcal{E}^\dagger \mathcal{E}|\Psi_\gamma, \bar{\Psi}_\beta\rangle.
	\end{aligned} 
\end{equation}
This means that the anyon overlap parameter $\langle {\alpha \bar{\beta}} | {\gamma \bar{\delta}} \rangle $ is invaraint under the exchange of the two copies in $\mathcal{H}/\bar{\mathcal{H}}$, i.e.,  $\langle {\alpha \bar{\beta}} | {\gamma \bar{\delta}} \rangle  =\langle {\gamma \bar{\beta}} | {\alpha \bar{\delta}} \rangle =\langle \alpha \bar{\delta} |{\gamma \bar{\beta} } \rangle  $.
To see that the EMD and partial-transpose symmetry together are inconsistent with any anyon condensed phase, we will first use the EMD symmetry to rule out all  possibilities except for the following two: (a)  both $e \bar{e}$ and $m \bar{m}$ are condensed. (b) both $e \bar{m}$ and $m \bar{e}$ are condensed.
We will later show that these two situations are impossible if one further insists on the partial-transpose symmetry.

Since $ \mathbb{U}_D = U_D \bar{U}_D$ exchanges charges and fluxes on $\mathcal{H}$ and $\bar{\mathcal{H}}$ simultaneously up to a lattice translation and $|\rho_0\rangle$ is the eigenstate of $ \mathbb{U}_D $, any anyon overlap parameters $\langle {\alpha \bar{\beta}} | {\gamma \bar{\delta}} \rangle $ without involving fermions are invariant under the exchange of charges and fluxes on $\mathcal{H}$ and $\bar{\mathcal{H}}$ simultaneously.
Taking $\langle I \bar{I}|m \bar{m} \rangle \propto \langle  \rho_0 |  \mathcal{E}^\dagger \mathcal{E}  w_m(\tilde{l}) \bar{w}_m (\tilde{l})|\rho_0 \rangle$ as an example, one finds 
\begin{equation}
	\label{Eq:mm_equal_ee}
	\begin{aligned}
		\langle & \rho_0 |  \mathcal{E}^\dagger \mathcal{E}  w_m(\tilde{l}) \bar{w}_m (\tilde{l})|\rho_0 \rangle  \\&  =  \langle \rho_0 | \mathbb{U}_D^\dagger  \mathbb{U}_D \mathcal{E}^\dagger \mathcal{E} \mathbb{U}_D^\dagger \mathbb{U}_D w_m (\tilde{l}) \bar{w}_m (\tilde{l}) \mathbb{U}_D^\dagger \mathbb{U}_D   |\rho_0 \rangle \\
		& =  \langle \rho_0 | (\mathbb{U}_D \mathcal{E}^\dagger \mathcal{E} \mathbb{U}_D^\dagger) (\mathbb{U}_D w_m (\tilde{l}) \bar{w}_m (\tilde{l})  \mathbb{U}_D^\dagger) |\rho_0 \rangle \\
		& =  \langle \rho_0 |  \mathcal{E}^\dagger \mathcal{E} w_e ({l} + 2\delta) \bar{w}_e ({l} + 2\delta)  |\rho_0 \rangle \\
		& =   \langle \rho_0 |  \mathcal{E}^\dagger \mathcal{E} w_e ({l} ) \bar{w}_e({l} )  |\rho_0 \rangle,
	\end{aligned}
\end{equation}
which implies $\langle I \bar{I} | m \bar{m}\rangle = \langle I \bar{I} | e \bar{e}\rangle   $.
In the final line, we have used the fact that both $\mathcal{E}$ and $|\rho_0\rangle$ are translationally invariant.
It is then straightforward to see that the weak EMD symmetry rules out the possibility of condensing self-bosons merely in $\mathcal{H}$ and $\bar{H}$.
For example, condensing $e$ implies $m$ is also condensed (since $\langle I \bar{I }| e \bar{I} \rangle = \langle I \bar{I}|m \bar{I}\rangle$ by EMD symmetry), which is in contradiction with the standard anyon condensation rule that the anyons ($m$)  with non-trivial mutual statistics with the condensed self-boson ($e$)  should be confined.
Therefore, if one requires $|\rho\rangle$  to encounter a transition through anyon condensation in the presence of weak EMD symmetry, only bound anyons in both $\mathcal{H}$ and $\bar{\mathcal{H}}$ can condense, which leaves us with three possibilities distinguished by their condensate generators: (a) condensation of $e \bar{e}$ and $m \bar{m}$, (b) condensation of $e \bar{m} $ and $m \bar{e}$, or (c) condensation only of $f \bar{f} = em \bar{e} \bar{m}$.
It turns out that the situation (c) can also be ruled out by the EMD symmetry,  as we now argue that condensing $f \bar{f}$ in fact implies that both $e \bar{e}$ and $m \bar{m}$ must also condense. To see this, let's
use Eq.\eqref{Eq:fermion_conjugate} to find how the $f\bar{f}$ creation operators $w_f(l) \bar{w}_f(l) = w_e(l) \bar{w}_e(l) w_m(\tilde{l}) \bar{w}_m(\tilde{l})$ transform under $\mathbb{U}_D$: $\mathbb{U}_D w_f (l) \bar{w}_f(l) \mathbb{U}_D^\dagger = [w_e(l+2 \delta) w_m(\tilde{l})] [\bar{w}_e(l+2 \delta)  \bar{w}_m(\tilde{l})] = w_{f,\pi} (l) \bar{w}_{f, \pi}$, where we denote  $w_{f,\pi} \equiv w_e(l+2 \delta) w_m(\tilde{l}) $ as the fermion creation operators rotated by angle $\pi$.
Therefore, the same argument in Eq.\eqref{Eq:mm_equal_ee} shows that the EMD symmetry requires $w_{f,\pi} (l) \bar{w}_{f, \pi}(l)$ to be condensed whenever  $w_f(l) \bar{w}_f(l)$ is condensed.
However, fusing $w_f(l) \bar{w}_f(l)$ and $w_{f,\pi} (l) \bar{w}_{f,\pi}(l)$ together yields $w_e(l) \bar{w}_e(l) w_e(l+ 2 \delta ) \bar{w}_e(l + 2 \delta)$, which corresponds to creating four $e \bar{e}$ anyons at the endpoints of $l$ and $l + 2\delta$. 
It is then obvious that repeatedly fusing $w_f \bar{w}_f$ and $w_{f,\pi}  \bar{w}_{f,\pi}$ corresponds to creating four $e \bar{e}$ anyons far apart.
Therefore, condensing $f \bar{f}$ along with the EMD symmetry implies that $e \bar{e}$ is also condensed. 
One can use the similar argument to show that $m \bar{m}$ is condensed as well. 

The remaining two possibilities that are compatible with the EMD symmetry are: (a) condensing $e \bar{e}$ and $m \bar{m}$, and (b) condensing $e \bar{m}$ and $m \bar{e}$.
We now use triangle (Cauchy-Schwarz) inequality to show that that both of these possibilities are also disallowed if one further imposes the aforementioned partial-transpose symmetry.
Let's assume  $|\rho\rangle$ encounters a transition through condensing $e \bar{e}$ and $m \bar{m}$. This implies that $\langle I \bar{I}|e \bar{e}\rangle = \langle I \bar{I}|m \bar{m}\rangle  $ saturates to a finite constant. Since $e$ and $\bar{e}$ anyons have a non-trivial mutual statistics with $m \bar{m}$, they must be confined, which implies $\langle e \bar{I}|e\bar{I} \rangle = \langle I \bar{e}| I \bar{e} \rangle = 0$.
On the other hand, the partial-transpose symmetry requires $\langle I \bar{I}|e \bar{e}\rangle = \langle I \bar{e}|e \bar{I}\rangle$ [due to Eq.\eqref{Eq:partial_TR}] to be a finite constant. 
However, a simple use of Cauchy-Schwarz inequality $\langle A|B\rangle \leq \sqrt{\langle A|A\rangle \langle B |B\rangle}$ yields 
\begin{equation}
	\label{Eq:cauchy_schwarz}
	\begin{aligned}
		\big(\langle \rho_0|&  \bar{w}_e \mathcal{E}^\dagger \big)  \big( \mathcal{E}w_e| \rho_0\rangle \big) \\
		& \leq \sqrt{\langle \rho_0| \bar{w}_e \mathcal{E}^\dagger \mathcal{E} \bar{w}_e | \rho_0\rangle\langle \rho_0| w_e \mathcal{E}^\dagger  \mathcal{E}w_e| \rho_0\rangle}.
	\end{aligned}
\end{equation}
After dividing $\langle \rho_0 |\mathcal{E}\mathcal{E}^\dagger|\rho_0 \rangle $ on both sides, Eq.\eqref{Eq:cauchy_schwarz} implies
$\langle I \bar{e}|e \bar{I}\rangle \leq \sqrt{\langle e \bar{I}|e\bar{I} \rangle  \langle I \bar{e}|I\bar{e} \rangle} = 0$, which is in contradiction with  $\langle I \bar{e}|e \bar{I}\rangle $ being a finite constant.
Therefore, the assumption of $|\rho\rangle$ encountering a phase transiton thorough condensing both $e\bar{e}$ and $m \bar{m}$ must be wrong.
One can similarly show that condensing $e \bar{m}$ and $m \bar{e}$ is disallowed: partial-transpose symmetry requires $\langle e \bar{I}| I \bar{m}\rangle = \langle I \bar{I} | e \bar{m} \rangle$ being a finite constant.
However, condensing  $e \bar{m}$ and $m \bar{e}$ implies confining $e$ and $\bar{m}$, i.e., $\langle e \bar{I} | e \bar{I} \rangle = \langle I \bar{m} | I \bar{m} \rangle  = 0$, which violates the Cauchy-Schwarz inequality $\langle e \bar{I}| I \bar{m}\rangle \leq \sqrt{\langle e \bar{I} | e \bar{I} \rangle \langle I \bar{m} | I \bar{m} \rangle }$.
As a result, we reach the conclustion that the C-J transformed channel $\mathcal{E}$ respects both EMD and partial-transpose symmetry, any transition out of the topologically ordered toric code phase must lie beyond the standard ayon condensation scheme. In the next two sections we consider channels that do indeed respect EMD and partial-transpose symmetries, and consider the fate of the toric code ground state and its close relative due to the action of such channels.

\section{Example 1: self-dual coherent errors in toric code}
\label{sec:coherent}
We now consider one simple example that satisfies the aforementioned conditions for a phase transition that can't be described within the standard anyon condensation scheme.
Specifically, we subject the toric code ground state to the composition of the following local channel on all edges $e$:
\begin{equation}
	\label{Eq:coherent_selfdual}
	\mathcal{E}_e[\rho] = (1-p) \rho + p \sigma_e \rho \sigma_e,\ \sigma_e  = \frac{Z_e +X_e}{\sqrt{2}}.
\end{equation}
Since $\sigma_e$ creates coherent superposition between states with different types of anyons, we will use the adjective `coherent' for the channel in Eq.\eqref{Eq:coherent_selfdual}.
After C-J map, $\mathcal{E} = \prod_e [(1-p) +p \sigma_e \bar{\sigma}_e]$ and one can easily show that ${U_\text{D} \bar{U}_\text{D} \mathcal{E} (U_\text{D} \bar{U}_\text{D})}^\dagger = \mathcal{E}$.
Besides, using the property $\sigma^T_e = \sigma_e$  and $\mathcal{E}_p^\dagger \mathcal{E}_p = \mathcal{E}_{p'}$ with $p' = 2p(1-p)$, one finds  $({\mathcal{E}}^\dagger\mathcal{E})^{T_{\mathcal{H}}} = ({\mathcal{E}}^\dagger\mathcal{E})^{{T}_{\bar{\mathcal{H}}}} = {\mathcal{E}}^\dagger\mathcal{E}$. 
Therefore, the transition for $|\rho\rangle$ out of the double topologically ordered state should lie beyond the anyon condensation scheme as discussed in the previous section.

In the following, we will map the transition induced by the channel in Eq.\eqref{Eq:coherent_selfdual} to several observables of the isotropic Ashkin-Teller (AT) model along the self-dual line, which explicitly shows that $|\rho\rangle$ encounters a  Berezinskii–Kosterlitz–Thouless (BKT) transition \cite{berezinskii1971destruction,Kosterlitz1973}.
Interestingly, this transition happens at the maximum error rate (i.e. $p=1/2$), and this is suggestive that the mixed state is topologically ordered up to $p = 1/2$. Later, in Sec.\ref{sec:nonfixed_coherent}, we will apply the channel in Eq.\ref{sec:nonfixed_coherent} to a \textit{non-fixed point} toric code which will allow us  to access an unconventional phase transition out of the topologically ordered phase into a mixed state with critical correlations that is stable over a whole range of parameters, as long as the EMD symmetry is preserved.

\subsection{Statistical mechanical mapping}
\label{sec:stat_mech}
The method of mapping the double state of toric code under decoherence to a statistical mechanical model follows closely to the idea in  Ref.\cite{zhu2019gapless}, where a pure-state transition for a single toric code was studied.
Since the overall scheme is fairly general and independent of the channel, we will first discuss the mapping without specifying the explicit form of the channel.
We will later focus on the channel in Eq.\eqref{Eq:coherent_selfdual} and show that the corresponding classical model is the AT model along the self-dual line.

To begin with, we make use of the duality \cite{kramers1941statistics,wegner1971duality,kubica2018ungauging} by writing toric code as the wave function  $|\Psi_0\rangle$  obtained from performing forced measurement of the 2d cluster state \cite{raussendorf2005long,aguado2008creation}. The Hilbert state of the cluster state consists of qubits both on the edges and vertices of a square lattice.
The cluster Hamiltonian takes the form
$H_{\textrm{cluster}} = \sum_v h_{v} + \sum_e h_{e}$, where
$h_v = - X_v A_v = -X_v \prod_{e \in v} Z_e$ and $h_e = - X_e Z_v Z_{v'}$ with $v$ and $v'$ denote the two vertices connected to the edge $e$.
It follows that the toric code wave function can be written as
\begin{equation}
	\label{Eq:toric_wavefunction}
	\begin{aligned}
		|\Psi_0\rangle & \propto \langle x_\mathbf{v} = 1| \Psi_{\text{cluster}} \rangle \\
		& \propto  \langle x_\mathbf{v} = 1 | \prod_e (I + X_e Z_v Z_{v'})|x_\mathbf{v} = 1, z_\mathbf{e} = 1\rangle \\
		& =  \sum_{x_\mathbf{e}} \sum_{z_v} \prod_e (1+ x_{e} z_v z_{v'}) |x_\mathbf{e}\rangle.
	\end{aligned}
\end{equation}
In the second line, we use $|\Psi_{\text{cluster}}\rangle = \prod_e (1+h_e)|x_\mathbf{v} = 1, z_\mathbf{e}= 1\rangle$ with $|x_\mathbf{v} = 1, z_\mathbf{e}= 1\rangle$ the product state satisfying $h_v |x_\mathbf{v} = 1, z_\mathbf{e}= 1 \rangle = 1, \forall e$. 
On the other hand, the final line can be obtained by inserting the complete basis $\{ |z_\mathbf{v}, x_\mathbf{e} \rangle\}$ between $(I + X_e Z_v Z_{v'})$ and $|x_\mathbf{v} = 1, z_\mathbf{e} = 1\rangle$.

Let's now investigate how Eq.\eqref{Eq:toric_wavefunction} assists our understanding of the effect of decoherence and the aforementioned anyon condensation and confinement parameters.
We first compute the normalization of $|\rho_0\rangle = |\Psi_0, \bar{\Psi}_0\rangle$.
Denoting $\langle \Psi_0|= \sum_{x_\mathbf{e}} \sum_{t_{\mathbf{v}}} \prod_e (1+ x_{e} t_v t_{v'}) \langle x_\mathbf{e}|$, one finds the wavefunction normalization $\langle \Psi_0 | \Psi_0\rangle = \sum_{z_\mathbf{v}, t_\mathbf{v}} \prod_e (1+z_v z_{v'} t_v t_{v'}) = \sum_{z_\mathbf{v}, t_\mathbf{v}} e^{\infty \sum_e z_v z_{v'} t_v t_{v'}}$.
It follows that
\begin{equation}
	\label{Eq:rho_0_rho_0}
	\langle \rho_0 | \rho_0\rangle = \sum_{z_\mathbf{v}, \bar{z}_\mathbf{v}, t_\mathbf{v}, \bar{t}_\mathbf{v} }  e^{\infty \sum_e (z_v z_{v'} t_v t_{v'} + \bar{z}_v \bar{z}_{v'} \bar{t}_v \bar{t}_{v'})}.
\end{equation}
Therefore, in the absence of decoherence, the Ising matter fields corresponding to the state $ |\Psi_0\rangle$, i.e., $z_\mathbf{v}$ and $t_\mathbf{v}$ are strongly coupled with each other, and similarly those originating from the state $|\bar{\Psi}_0\rangle$, i.e., $\bar{z}_\mathbf{v}$) and  $\bar{t}_\mathbf{v}$) are also strongly coupled.
On the other hand, when the decoherence is introduced, the normalization of the decohered double state $|\rho\rangle = \mathcal{E}|\rho_0\rangle$ takes the form
\begin{equation}
	\begin{aligned}
		\langle  \rho|\rho\rangle & 
		= \sum_{z_\mathbf{v}, \bar{z}_\mathbf{v}, t_\mathbf{v}, \bar{t}_\mathbf{v} }  e^{-H(z_\mathbf{v}, \bar{z}_\mathbf{v}, t_\mathbf{v}, \bar{t}_\mathbf{v})}, 
	\end{aligned}
\end{equation}
where
\begin{equation}
	\label{Eq:H_nonfixedpoint}
	\begin{aligned}
		& e^{-H(z_\mathbf{v}, \bar{z}_\mathbf{v}, t_\mathbf{v}, \bar{t}_\mathbf{v})}  = \sum_{\substack{x_\mathbf{e}, \bar{x}_\mathbf{e}, \\
				x'_\mathbf{e}, \bar{x}'_\mathbf{e} }}
		\Big[ \langle x'_\mathbf{e}, \bar{x}'_\mathbf{e} |{\mathcal{E}}^\dagger {\mathcal{E}}|x_\mathbf{e}, \bar{x}_\mathbf{e}\rangle \times \\
		& \prod_e (1 + x_e z_v z_{v'}) (1 + \bar{x}_e \bar{z}_v \bar{z}_{v'}) (1 + x'_e t_v t_{v'}) (1 + \bar{x}'_e \bar{t}_v \bar{t}_{v'}) 
		\Big].
	\end{aligned}
\end{equation}
While we haven't specified the explicit form of the operator $\mathcal{E}$, there is a constraint that $\mathcal{E}$ is a finite-depth local operator since $\mathcal{E}[\cdot]$ is a finite-depth composition of local channels.
This implies that $H$ should be a local 2d Hamiltonian, as the four Ising matter fields only interact with one another through $\langle x'_\mathbf{e}, \bar{x}'_\mathbf{e} |{\mathcal{E}}^\dagger {\mathcal{E}}|x_\mathbf{e}, \bar{x}_\mathbf{e}\rangle $ in Eq.\eqref{Eq:H_nonfixedpoint}.
Another essential feature of $H(z_\mathbf{v}, \bar{z}_\mathbf{v}, t_\mathbf{v}, \bar{t}_\mathbf{v})$ is that it will be invariant under the individual global spin-flip operation $s_\mathbf{v} \rightarrow -s_\mathbf{v}$ for any $s_\mathbf{v} = z_\mathbf{v}, \bar{z}_\mathbf{v}, t_\mathbf{v}, \bar{t}_\mathbf{v}$.
This can be seen by noting that each term in the summation of Eq.\eqref{Eq:H_nonfixedpoint} is invariant under the individual global spin-flip operation.
Therefore, it is natural to expect that different quantum phases of $|\rho\rangle$ are mapped to different classical phases of $H$ depending on what symmetries are spontaneously broken.
For example, from Eq.\eqref{Eq:rho_0_rho_0}, one finds that the initial double toric code $|\Psi_0, \bar{\Psi}_0 \rangle$ is in the double partial ordered phase characterized by $\langle z_i t_i z_j t_j \rangle = \langle \bar{z}_i \bar{t}_i \bar{z}_j \bar{t}_j \rangle$  saturating to a non-zero constant as $|i - j |\rightarrow \infty$ while $\langle z_i z_j \rangle = \langle t_i t_j \rangle =\langle \bar{z}_i \bar{z}_j \rangle =\langle \bar{t}_i \bar{t}_j \rangle $ decay exponentially as a function of $|i-j|$.
Remarkably, these order parameters precisely corrrespond to the $e$ condensation and confinement parameters \textit{no matter} what the operator $\mathcal{E}$ is (see Appendix \ref{sec:app_anyon_mapping} for details): 
\begin{equation}
	\label{Eq:EOP_mapping}
	\begin{aligned}
		\langle I \bar{I} |e \bar{I}\rangle & = \lim_{| i - j| \rightarrow \infty} \langle z_i z_j \rangle,\\ 
		\langle e \bar{I} |e \bar{I}\rangle & = \lim_{| i - j| \rightarrow \infty} \langle z_i t_i z_j t_j\rangle. \\
	\end{aligned}
\end{equation}
One can similarly show that the $e \bar{e}$ condensation parameter $	\langle I \bar{I} |e \bar{e}\rangle$ is mapped to $ \lim_{| i - j| \rightarrow \infty} \langle z_i\bar{z}_i z_j \bar{z}_j\rangle$ and also vanishes for the intial double toric code.
On the other hand, the $m \bar{m}$ condensation parameter is mapped to the following disorder parameter:
\begin{equation}
	\label{Eq:EOP_mapping2}
	\begin{aligned}
		\langle I \bar{I} |m \bar{m}\rangle & =  \lim_{|\tilde{i} -\tilde{j}| \rightarrow  \infty }\frac{\sum_{z_\mathbf{v}, \bar{z}_\mathbf{v}, t_\mathbf{v}, \bar{t}_\mathbf{v} } e^{- H_{ z_{\tilde{l}}\bar{z}_{\tilde{l}}}(z_\mathbf{v}, \bar{z}_\mathbf{v}, t_\mathbf{v}, \bar{t}_\mathbf{v})  } }{\sum_{z_\mathbf{v}, \bar{z}_\mathbf{v}, t_\mathbf{v}, \bar{t}_\mathbf{v}} e^{- H_{} (z_\mathbf{v}, \bar{z}_\mathbf{v}, t_\mathbf{v}, \bar{t}_\mathbf{v})}}  \\& \equiv \lim_{|\tilde{i} - \tilde{j}| \rightarrow \infty} \langle \mu^{z}_{\tilde{i}} \mu^{\bar{z}}_{\tilde{i}} \mu^{z}_{\tilde{j}} \mu^{\bar{z}}_{\tilde{j}} \rangle.
	\end{aligned}
\end{equation}
Here $\tilde{l}$ denotes any string on the dual lattice with the endpoints located at sites $\tilde{i}$ and $\tilde{j}$, and 
\begin{equation}
	\begin{aligned}
		&  e^{- H_{ z_{\tilde{l}}\bar{z}_{\tilde{l}}}(z_\mathbf{v}, \bar{z}_\mathbf{v}, t_\mathbf{v}, \bar{t}_\mathbf{v})  }  = \sum_{\substack{x_\mathbf{e}, \bar{x}_\mathbf{e}, \\
				x'_\mathbf{e}, \bar{x}'_\mathbf{e} }}
		\Big[ \langle x'_\mathbf{e}, \bar{x}'_\mathbf{e} |{\mathcal{E}}^\dagger {\mathcal{E}}|x_\mathbf{e}, \bar{x}_\mathbf{e}\rangle \times \\
		\prod_e & (1 + \eta_e x_e z_v z_{v'}) (1 + \eta_e \bar{x}_e \bar{z}_v \bar{z}_{v'}) (1 + x'_e t_v t_{v'}) (1 + \bar{x}'_e \bar{t}_v \bar{t}_{v'}) 
		\Big]
	\end{aligned}
\end{equation}
with $ \eta_{e}= -1(+1)$ if $e \in \tilde{l} (\ni \tilde{l})$.
One  similarly finds that the $m$ confinement parameter $\langle m \bar{I} | m \bar{I}\rangle =\lim_{|\tilde{i} - \tilde{j}| \rightarrow \infty} \langle \mu^{z}_{\tilde{i}} \mu^{t}_{\tilde{i}} \mu^{z}_{\tilde{j}} \mu^{t}_{\tilde{j}} \rangle$. Here $\langle \mu^{z}_{\tilde{i}} \mu^{t}_{\tilde{i}} \mu^{z}_{\tilde{j}} \mu^{t}_{\tilde{j}} \rangle$ is defined analogously to $\langle  \mu^{z}_{\tilde{i}} \mu^{\bar{z}}_{\tilde{i}} \mu^{z}_{\tilde{j}} \mu^{\bar{z}}_{\tilde{j}} \rangle$ (see Eq.\eqref{Eq:EOP_mapping2}).
In fact, we can identify  anyon overlap parameters $\langle {\alpha \bar{\beta}} | {\gamma \bar{\delta}} \rangle \equiv \langle \rho_{\alpha \bar{\beta}} | \rho_{\gamma \bar{\delta}} \rangle / \langle \rho | \rho\rangle$ as certain combinations of order and disorder parameters in the statistical model (see Appendix \ref{sec:app_anyon_mapping}).

Another interesting feature of the mapping is that the anyon condensation rules now correspond to the constraints on order and disorder parameters of any $\mathbb{Z}_2$ symmetric system with finite correlation length \cite{levin2020constraints}.
For example, one finds $\langle I \bar{I} |e \bar{e}\rangle \propto \langle z_i\bar{z}_i z_j \bar{z}_j\rangle $ detects the spontaneous breaking of any symmetries that $z_i \bar{z}_j$ is charged under while $	\langle m \bar{I}| m \bar{I} \rangle \propto  \langle \mu^{z}_{\tilde{i}} \mu^{{t}}_{\tilde{i}} \mu^{z}_{\tilde{j}} \mu^{{t}}_{\tilde{j}} \rangle$ detects the disorder of $z_\mathbf{v} t_\mathbf{v}$.
The anyon condensation rule that the condensation of $e \bar{e}$ implies the confinement of $m$ 
then corresponds to the constraint that the order and disorder parameters of the spin-flip operation $(z_\mathbf{v}, t_\mathbf{v}) \rightarrow -(z_\mathbf{v}, t_\mathbf{v})$ cannot be nonzero at the same time for any finite correlation length system \cite{levin2020constraints}, and thus $\lim_{| \tilde{l}| \rightarrow  \infty }  \langle \mu^{z}_{\tilde{i}} \mu^{{t}}_{\tilde{i}} \mu^{z}_{\tilde{j}} \mu^{{t}}_{\tilde{j}}  \rangle$ must vanish when $\lim_{| i - j| \rightarrow \infty} \langle z_i\bar{z}_i z_j \bar{z}_j\rangle $ is a finite constant. This statement requires an assumption that the lowest  eigenvalue of the corresponding transfer matrix within the even subspace is non-degenerate, see Ref.\cite{o2018lattice} for an example where this assumption is violated.
Given this connection, it's illuminating to investigate how the weak EMD and partial-transpose symmetry imply $|\rho\rangle$ must induce transition (if any) beyond anyon condensation scheme using the constraints on order and disorder parameters.
Let's assume that $e \bar{e}$ and $m \bar{m}$ condense simultaneously, i.e., $\lim_{|i-j| \rightarrow \infty }\langle z_i \bar{z}_i z_j \bar{z}_j\rangle =\lim_{|\tilde{i} - \tilde{j}| \rightarrow \infty } \langle \mu^{z}_i \mu^{\bar{z}}_i \mu^{z}_j \mu^{\bar{z}}_j \rangle $ is a finite constant (the situation of condensing $e \bar{m}$ and $m \bar{e}$ can be analyzed using similar arguments).
On the other hand, the partial-transpose symmetry means $\lim_{|i-j| \rightarrow \infty }\langle z_i \bar{t}_i z_j \bar{t}_j\rangle = \lim_{|i-j| \rightarrow \infty }\langle z_i \bar{z}_i z_j \bar{z}_j\rangle$ is also nonzero.
However, since $\langle \mu^{z}_i \mu^{\bar{z}}_i \mu^{z}_j \mu^{\bar{z}}_j \rangle$ is the disorder parameter of the symmetry transformation $(z_\mathbf{v}, \bar{z}_\mathbf{v} ) \rightarrow - (z_\mathbf{v}, \bar{z}_\mathbf{v} )$ and  $z_i \bar{t}_i $ is charged under this symmetry, $\lim_{|i-j| \rightarrow \infty }\langle z_i \bar{t}_i z_j \bar{t}_j\rangle $ and $\lim_{|\tilde{i} - \tilde{j}| \rightarrow \infty } \langle \mu^{z}_i \mu^{\bar{z}}_i \mu^{z}_j \mu^{\bar{z}}_j \rangle$ cannot be nonzero at the same time for any system with finite correlation length.
We therefore reach a contradiction with the initial assumption that $e \bar{e}$ and $m \bar{m}$ are condensed at the same time.

Having discussed the general mapping, we now restrict ourselves to the self-dual coherent channel in Eq.\eqref{Eq:coherent_selfdual} and show the it is mapped to the AT model along the self-dual line.
We calculate the normalization of the double state in Appendix \ref{sec:app_coherent}, and find that
\begin{equation}
	\begin{aligned}
		\label{Eq:doubleAT_dual}
		\langle \rho|\rho\rangle & =  \sum_{z_\mathbf{v}, \bar{z}_\mathbf{v}, t_\mathbf{v} }e^{-H(z_\mathbf{v}, \bar{z}_\mathbf{v}, t_\mathbf{v}) }, \\
		-H & = K \sum_e (z_v z_{v'} \bar{z}_v \bar{z}_{v'} + t_v t_{v'} \bar{z}_v \bar{z}_{v'} ) + K_4\sum_e z_v z_{v'} t_v t_{v'}, \\
	\end{aligned}
\end{equation}
where $K$ and $K_4$ are determined by $p$ through
\begin{equation}
	\label{Eq:KK4_h}
	\begin{aligned}
		\tanh(K)  = \frac{2- \sqrt{4-{\lambda}^2}}{\lambda},\
		\tanh(K_4)  = \frac{2- \lambda \sqrt{4-{\lambda}^2}}{2-{\lambda}^2},
	\end{aligned}
\end{equation}
with $\lambda = 2p(1-p)/ (1-2p+2p^2)$.
In particular, $p = 0$ corresponds to $(K,K_4) = (0, \infty)$ while $p = 1/2$ corresponds to $K = K_4 = \tanh^{-1}(2-\sqrt{3}) \approx 0.275$.
Therefore, increasing decoherence effectively decreases the intra-copy interaction, while increasing the inter-copy interaction.
Note that we have used the fact that $ z_v z_{v'}  \bar{z}_v \bar{z}_{v'}  t_v t_{v'}\bar{t}_{v}\bar{t}_{v'} =1, \forall e$ to eliminate $\bar{t}_\mathbf{v}$ in Eq.\eqref{Eq:doubleAT_dual} (see Appendix \ref{sec:app_coherent} for details).
It is also worth mentioning that the coupling stengths of $z_v z_{v'} \bar{z}_v \bar{z}_{v'}$ and $t_v t_{v'} \bar{z}_v \bar{z}_{v'}$ being equal is precisely due to the partial-transpose symmetry of $\mathcal{E}^\dagger \mathcal{E}$.
To show the equivalence between $\langle \rho| \rho\rangle$ and the partition function of the isotropic Ashkin-Teller model, we define  new variables $s_v = z_v \bar{z}_v,\ \tau_v = t_v \bar{z}_v$ and write Eq.\eqref{Eq:doubleAT_dual}  as
\begin{equation}
	\label{Eq:AT_reduce}
	-H = K \sum_{e} (s_v s_{v'}  + \tau_v \tau_{v'}) + K_4 \sum_{e}  s_v s_{v'}  \tau_v \tau_{v'},
\end{equation}
which is the desired form of the isotropic AT model.
Besides, Eq.\eqref{Eq:KK4_h} implies $ K$ and $K_4$ are related to each other through $e^{-2K_4} = \sinh(2K) $, which is the self-duality condition of the AT model.
This is not a coincidence, as the self-duality condition essentially corresponds to the average EMD duality.
To make their connection more precise, we rewrite the $e \bar{e}$ condensation parameter $\langle I \bar{I} |e \bar{e}\rangle$ and the $e$ confinement parameter $\langle e \bar{I}|e \bar{I}\rangle$ in terms of the new variables, and they take the form
\begin{equation}
	\label{Eq:EOP_mapping}
	\begin{aligned}
		\langle I \bar{I} |e \bar{e}\rangle & = \lim_{| i - j| \rightarrow \infty} \langle z_i\bar{z}_i z_j \bar{z}_j\rangle = \lim_{| i - j| \rightarrow \infty} \langle s_i s_j\rangle, \\
		\langle e \bar{I} |e \bar{I}\rangle & = \lim_{| i - j| \rightarrow \infty} \langle z_i t_i z_j t_j\rangle = \lim_{| i - j| \rightarrow \infty} \langle s_i \tau_i s_j \tau_j \rangle. \\
	\end{aligned}
\end{equation}
Similarly, one finds 
\begin{equation}
	\begin{aligned}
		\langle I \bar{I} | m \bar{m}\rangle
		& = 
		\lim_{|\tilde{i} - \tilde{j}| \rightarrow \infty} \langle \mu^{z}_{\tilde{i}} \mu^{\bar{z}}_{\tilde{i}} \mu^{z}_{\tilde{j}} \mu^{\bar{z}}_{\tilde{j}} \rangle \\
		& =  \lim_{|\tilde{i} - \tilde{j}| \rightarrow \infty} \langle \mu^{\tau}_{\tilde{i}}  \mu^{\tau}_{\tilde{j}} \rangle  =  \lim_{|\tilde{i} - \tilde{j}| \rightarrow \infty} \langle \mu^{s}_{\tilde{i}}  \mu^{s}_{\tilde{j}} \rangle,\\
		\langle m \bar{I} | m \bar{I}\rangle & =\lim_{|\tilde{i} - \tilde{j}| \rightarrow \infty} \langle \mu^{z}_{\tilde{i}} \mu^{{t}}_{\tilde{i}} \mu^{z}_{\tilde{j}} \mu^{{t}}_{\tilde{j}} \rangle \\
		& = \lim_{|\tilde{i} - \tilde{j}| \rightarrow \infty} \langle \mu^{s}_{\tilde{i}} \mu^{\tau}_{\tilde{i}} \mu^{s}_{\tilde{j}} \mu^{\tau}_{\tilde{j}} \rangle.
	\end{aligned}
\end{equation}
It is then obvious that requiring $\langle I \bar{I}| e \bar{e} \rangle =\langle I \bar{I}| m \bar{m} \rangle $ and $\langle e \bar{I}| e \bar{I} \rangle =\langle m \bar{I}| m \bar{I} \rangle $ forces the classical model to lie along the self-dual line.

\begin{figure}
	\centering
	\includegraphics[width=\linewidth]{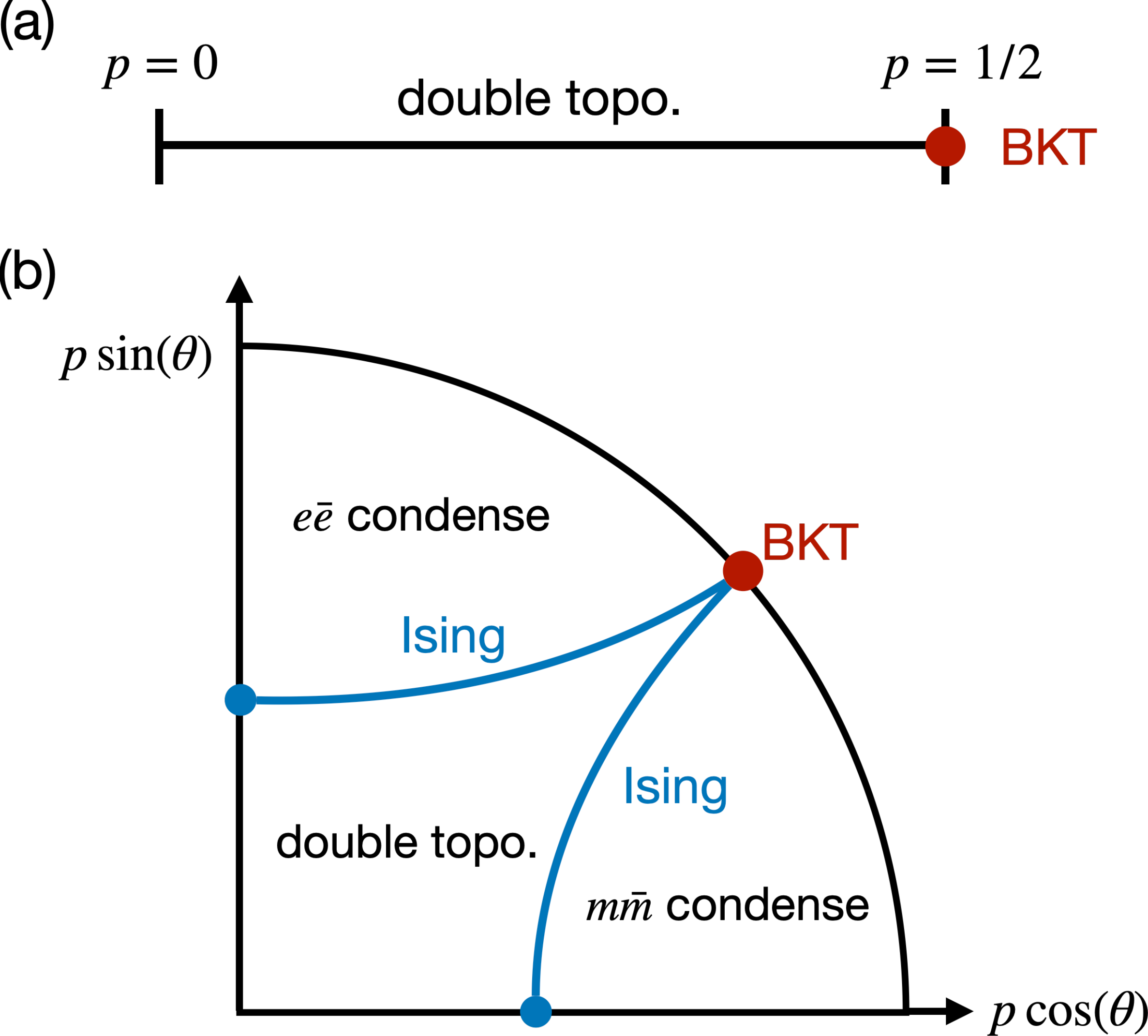}
	\caption{(a) The phase diagram for the double state of toric code subjected to the self-dual coherent channel with Kraus operator $\sigma_e = (Z_e + X_e)/\sqrt{2}$. The BKT transition happens at the maximum error rate $p=1/2$, and before that $|\rho\rangle$ possesses double topologically ordered. 
		(b) The phase diagram of the toric code subjected to a more general coherent channel with with Kraus operator $\sigma_e = \cos(\theta)Z_e + \sin(\theta)X_e$. 
	}
	\label{Fig:coherent_phase}
\end{figure}

The exact mapping allows us to borrow the knowledge of the AT model \cite{baxter2016exactly} to understand the fate of the decohered mixed state [see Fig.\ref{Fig:coherent_phase}(a)].
Along the self-dual line of the isotropic AT model and when $K_4 > K (p<1/2)$, it belongs to the partial ordered phase with $\langle s_j\rangle = \langle \tau_j\rangle = 0$ while $\langle s_j \tau_j \rangle \neq 0$.
Therefore, $ \langle s_i s_j \rangle$ decays exponentially as a function of $|i-j|$ while $\lim_{|i-j| \rightarrow  \infty } \langle s_i \tau_i s_j \tau_j\rangle$ saturates to a finite constant.
This implies $e \bar{e}$ is not condensed and $e$ is deconfined, consistent with the double topologically ordered state.
On the other hand, when $K = K_4  = \tanh^{-1}(2-\sqrt{3}) $(corresponding to $p = 1/2$), the classical model hits the Berezinskii-Kosterlitz-Thouless (BKT) transition point and then enters into a critical line with continuously varying exponents for  $K\geq K_4$. However, for the decohered state, $K > K_4$ is not possible since $p = 1/2$ is maximum allowed error rate. In Sec.\ref{sec:nonfixed_coherent} we will access the BKT regime by starting with a perturbed  toric code.

\subsection{Convex decomposition at the maximum error rate}
The above mapping of the double state normalization to the AT model shows that the double topological order of $|\rho_0\rangle$ is stable as long as the error rate is less than the maximum value of $p = 1/2$. This is suggestive that perhaps the decohered mixed state $\rho$ is long-range entangled  for $p < 1/2$. To establish this, one would need to calculate a true measure of mixed-state entanglement such as entanglement negativity, or perhaps more ambitiously, show that the mixed state is not separable for $p < 1/2$, i.e., there exists no convex decomposition of the mixed state in terms of SRE states for $p < 1/2$. Here we would content ourselves with studying the separability aspects just at $p = 1/2$.
When $p = 1/2$, $\mathcal{E} \propto \prod_e (I + \sigma_e \bar{\sigma}_e)$ is a projector to the even sector of $ \sigma_e \bar{\sigma}_e$ and hence is non-invertible (i.e. has a non-trivial kernel).
Therefore, as mentioned in Sec.\ref{sec:condition}, the power-law decay of several anyon condensation parameters \textit{does not} guarantee that  $|\rho\rangle$ has long-range correlations.
Furthermore, even if the double state $|\rho\rangle$ has long-range correlations, they may  originate from classical correlations of the original density matrix $\rho$ and do not necessarily imply long-range quantum entanglement for  $\rho$ itself.

We now explictly show that $\rho$ at $p = 1/2$ can be written as a convex sum of product states, which guarantees that $\rho$ can at most have long-range classical correlations.
This can be done by noting that the Kraus operators for a given channel are non-unique, and the self-dual coherent channel in Eq.\eqref{Eq:coherent_selfdual} can be equivalently written as $\mathcal{E}_e[\rho] = K_{e,+} \rho K_{e,+} + K_{e,-}\rho K_{e,-}$, where the Kraus operators $K_{e, \pm}= (\sqrt{1-p}I \pm \sqrt{p} \sigma_e )/\sqrt{2}$.
Given such a representation, one can express  $\rho$ as the following convex decomposition:
\begin{equation}
	\label{Eq:nonoptimal}
	\begin{aligned}
		\rho & = \sum_{m_\mathbf{e} } \Big( \prod_e K_{e,m_e} |\Psi_0\rangle \Big) \Big( \langle \Psi_0| \prod_e K_{e,m_e} \Big)  \\
		& = \sum_{m_\mathbf{e}} |\psi_{m_\mathbf{e}}\rangle \langle \psi_{m_\mathbf{e}}| = \sum_{m_\mathbf{e}} P_{m_\mathbf{e}} |\tilde{\psi}_{m_\mathbf{e}}\rangle \langle \tilde{\psi}_{m_\mathbf{e}}|.
	\end{aligned}
\end{equation}
Here $m_\mathbf{e} = \{ m_e = \pm 1\}$, $|\psi_{m_\mathbf{e}}\rangle = \prod_e K_{e,m_e} |\Psi_0\rangle$ is the non-normalized wave function, $P_{m_\mathbf{e}} = \langle \psi_{m_\mathbf{e}} |  \psi_{m_\mathbf{e}}\rangle$ is the probability for $|\psi_{m_\mathbf{e}}\rangle$ to occur, and $|\tilde{\psi}_{m_\mathbf{e}}\rangle$ is the normalized version of $|\psi_{m_\mathbf{e}}\rangle$.
Since $K_{e,\pm}$ becomes a projector to the $\pm$ sector of $\sigma_e$ at $p = 1/2$, one finds $\sigma_e |\tilde{\psi}_{m_\mathbf{e}}\rangle = m_e |\tilde{\psi}_{m_\mathbf{e}}\rangle$, and thus $|\tilde{\psi}_{m_\mathbf{e}}\rangle $ is a product state in the $\sigma_e$ basis. 

It is also worth mentioning that the probability for each $|\psi_{m_\mathbf{e}}\rangle$ to occur at a given error rate $p$ takes the following interesting form:
\begin{equation}
	\begin{aligned}
		& P_{m_\mathbf{e}}= \\
		& \prod_e \Big[ 1  + \frac{m_e \mu }{\sqrt{2} + m_e}(z_v z_{v'}+t_v t_{v'})+ \frac{\sqrt{2} - m_e \mu }{\sqrt{2} + m_e \mu }z_v z_{v'}t_v t_{v'} \Big],
	\end{aligned}
\end{equation}
where $\mu = 2\sqrt{p(1-p)}$.
At $p= 1/2$,  one can use $\langle \tilde{\psi}_{m_\mathbf{e}'} | \tilde{\psi}_{m_\mathbf{e}}\rangle = \delta_{m_\mathbf{e}', m_\mathbf{e}}$ to show that $\tr(\rho^2) = \sum_{m_\mathbf{e}} P^2_{m_\mathbf{e}}$, which is the  \textit{classical} 2nd Renyi entropy of the convex decomposition.
Therefore, the density matrix at $p = 1/2$ possesses classical long-range correlations characterized by the classical 2nd Renyi entropy.

\subsection{More general coherent-error channel}

Let us briefly comment on the fate of toric code under a more general coherent channel with Kraus operator $\sigma_e(\theta) = \cos(\theta)Z_e + \sin(\theta)X_e$ which generically does not respect the EMD symmetry.
We find that $\langle \rho|\rho\rangle$ is now mapped to the general isotropic AT model, with $K$ and $K_4$ related to $\lambda$ and $\theta$ through (see Appendix \ref{sec:app_coherent} for details)
\begin{equation}
	\begin{aligned}
		\frac{\tanh(K) [1+\tanh(K_4)] }{1+\tanh^2(K) \tanh(K_4)} & =   \frac{\lambda \sin^2(\theta)}{1+\lambda \cos^2(\theta)} , \\
		\frac{\tanh^2(K) +\tanh(K_4)}{1+\tanh^2(K) \tanh(K_4)} & =  \frac{1- \lambda\cos^2(\theta)}{1+ \lambda \cos^2(\theta) } .
	\end{aligned}
\end{equation}
One can now use our knowledge about the phase diagram of the general isotropic ferromagnetic AT model to understand the phase diagram of the double state $|\rho\rangle$ subjected to the channel with Kraus operator $\sigma_e(\theta)$.
In particular, the ferromagnetic AT model has three phases \cite{baxter2016exactly}:

\noindent
(i) The partial ordered phase with $\langle s_j\rangle = \langle \tau_j\rangle = 0$ while $\langle s_j \tau_j \rangle \neq 0$.
In this regime, $\lim_{|i-j| \rightarrow  \infty } \langle s_i s_j\rangle = 0$ and thus $\langle I \bar{I} |e \bar{e} \rangle = 0$.
On the other hand, the ordering of $s_\mathbf{v} \tau_\mathbf{v}$ implies that the free energy cost of creating the domain wall for $s_\mathbf{v} \tau_\mathbf{v}$ must diverge with $|\tilde{l}|$.
As a result, $\lim_{| \tilde{l}| \rightarrow  \infty } \langle  \mu^{s}_{\tilde{i}} \mu^{s}_{\tilde{j}} \rangle =0$ and thus $\langle I \bar{I} | m \bar{m}\rangle = 0$.
One can similarly show that both $\langle e \bar{I} |e \bar{I} \rangle $ and $\langle m \bar{I} |m \bar{I}\rangle$ are finite.
Therefore, $|\rho\rangle$ is a double $\mathbb{Z}_2$ topologically ordered state, suggesting that the quantum memory may be retained for the mixed state $\rho$.

\noindent
(ii) The Baxter (ordered) phase with $\langle s_j\rangle = \langle \tau_j\rangle \neq 0$ and $\langle s_j \tau_j \rangle \neq 0$.
In this regime, $\lim_{|i-j| \rightarrow  \infty } \langle s_i s_j\rangle $ is a finite constant and thus $ \langle I \bar{I} |e \bar{e} \rangle $ is finite.
Similar to the partial ordered phase, the ordering of $\langle s_j\sigma_j \rangle$ implies $\langle I \bar{I} | m \bar{m}\rangle = 0$.
One can also show that both $\langle e \bar{I} |e \bar{I} \rangle $ and $\langle m \bar{I} |m \bar{I}\rangle$ are finite using similar arguments.
Therefore, $|\rho\rangle$  only has a single topological order due to the condensation of $e \bar{e}$.
This suggests that the mixed state $\rho$ only possesses classical memory.

\noindent
(iii) The paramagnetic phase with $\langle s_j\rangle = \langle \tau_j\rangle = 0$ and $\langle s_j \tau_j \rangle = 0$.
Due to the disordering of $\langle s_j\tau_j \rangle$, the free energy cost of creating the domain wall for $s_\mathbf{v} \tau_\mathbf{v}$ along the path $\tilde{l}$ now becomes finite independent of the system size.
Therefore, $\langle I \bar{I}|m \bar{m} \rangle $ saturates to a finite value.
On the other hand, $\langle I \bar{I}| e \bar{e}\rangle = 0$ similar to the partial ordered phase.
As a result, $|\rho\rangle$  only has a single topological order due to the condensation of $m \bar{m}$, suggesting that the mixed state $\rho$ only possesses classical memory.

We summarize these observations schematically in the phase diagram shown in Fig.\ref{Fig:coherent_phase}(b).

\section{Example 2: self-dual coherent errors in a perturbed toric code}
\label{sec:nonfixed_coherent}

\begin{figure*}
	\centering
	\includegraphics[width=\linewidth]{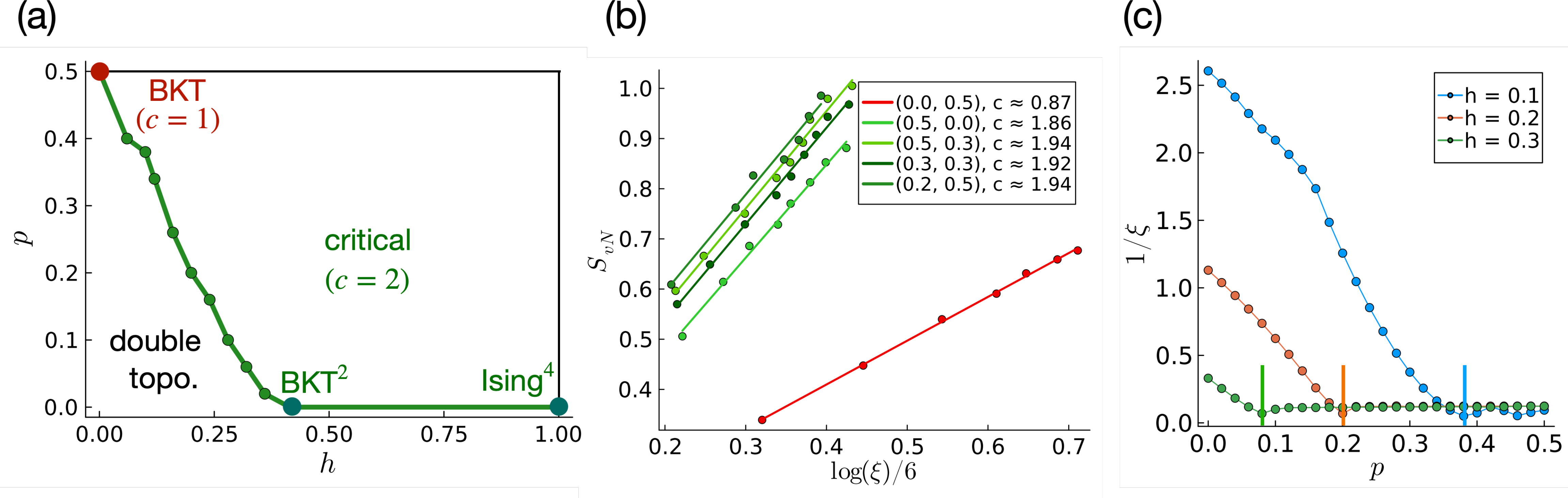}
	\caption{	(a) Phase diagram for the double state of the perturbed toric code (Eq. \eqref{Eq:selfdual_nonfixed}) subjected to the self-dual coherent channel. 
		Here the locations of BKT, $\text{BKT}^2$, and $\text{Ising}^4$ are known from exact mappings while the other green dots  are determined by calculations similar to (c).
		(b) Finite-entanglement scaling of the approximate MPS at different locations.
		The legend shows the numerically extracted central charges at various locations in the $(h,p)$-plane.
		(c) Inverse of the correlation length $1 /\xi$ as a function of the error rate $p$ along the lines $h = 0.1$, $0.2$, and $0.3$.
		The solid vertical lines mark the positions where $\xi$ starts to diverge.
	}
	\label{Fig:nonfixed_phase}
\end{figure*}

In the previous section we showed the double topologically ordered toric code ground state is stable under the self-dual coherent channel [Eq.\eqref{Eq:coherent_selfdual}] as long as $p<1/2$.
One may wonder whether this statement is true for any $\mathbb{Z}_2$ topologically ordered ground state subjected to self-dual coherent channel.
To make progress on this question, we consider applying the same channel as in the previous section to a perturbed toric code that has non-zero correlation length, and numerically study the corresponding statistical mechanics model. We find that the transition induced by the quantum channel can indeed happen below the maximum error rate, and as the topological order is destroyed, the system enters a critical mixed-state phase protected by the EMD symmetry (at least based on the quantities related to $\tr(\rho^2)$). 

We start with the following pure state corresponding to a toric code ground state $|\Psi_0\rangle$ perturbed by a non-unitary self-dual operator \cite{zhu2019gapless}: 
\begin{equation}
	\label{Eq:selfdual_nonfixed}
	|\Psi(h)\rangle = \mathcal{N}|\Psi_0\rangle = \prod_e (I + h \sigma_e)   |\Psi_0\rangle,
\end{equation}
where $h \in [0,1]$ and we again denote $\sigma_e = (X_e + Z_e)/\sqrt{2}$. 
We emphasize that since $\mathcal{N}$ is a product of single-site non-unitary operators, it acts as a temporal defect when considering the normalization of $|\Psi(h)\rangle$ in the path intergral representation.
Therefore, $\mathcal{N}$ can only induce a (2+0)-D transition instead of the generic (2+1)-D transition encountered when tuning parameters in a Hamiltonian.
In particular, it has been shown in 
Ref.\cite{zhu2019gapless}  that $|\Psi(h)\rangle $ remains topologically ordered when $h<h_c = \sqrt{2}-1 \approx 0.414$ while for $h>h_c$, it becomes a novel gapless state with continuously varying exponents. The correlation functions of the anyon string operators in this critical regime can be related to the conformal field theory corresponding to a free, massless, complex scalar. $h = 1$ corresponds to two decoupled critical Ising models, while $ h = h_c = \sqrt{2}-1$ corresponds to the BKT point at which an operator becomes relevant and opens a mass gap resulting in the topologically ordered phase for $h < h_c$.

Next, we subject $|\Psi(h)\rangle$ to  the self-dual coherent channel in Eq.\eqref{Eq:coherent_selfdual}. The corresponding double state can be written as $|\rho\rangle =\mathcal{E} (\mathcal{N} \otimes \bar{\mathcal{N}}) |\Psi_0, \bar{\Psi}_0\rangle$.
This amounts to evolving the double toric code $|\Psi_0, \bar{\Psi}_0\rangle$ through the non-unitary operator $\tilde{\mathcal{E}} = \mathcal{E} (\mathcal{N} \otimes \bar{\mathcal{N}})$.
Similar to the statistical mechanical mapping in Sec.\ref{sec:stat_mech}, we can write $\langle  \rho|\rho\rangle = \langle \Psi_0, \bar{\Psi}_0| \tilde{\mathcal{E}}^\dagger \tilde{\mathcal{E}} | \Psi_0, \bar{\Psi}_0\rangle$ as the partition function of 4-flavored Ising model that takes the form in Eq.\eqref{Eq:H_nonfixedpoint}.
One can also define the corrupted anyon state $|\rho_{\alpha \bar{\beta}} \rangle = \tilde{\mathcal{E}} |\rho_{\alpha \bar{\beta},0} \rangle = \tilde{\mathcal{E}}  w_\alpha(l) \bar{w}_\beta(l) |\rho_0\rangle$ and evaluate the anyon overlap parameters $\langle {\alpha \bar{\beta}} | {\gamma \bar{\delta}} \rangle \equiv \langle \rho_{\alpha \bar{\beta}} | \rho_{\gamma \bar{\delta}} \rangle / \langle \rho | \rho\rangle$.
They again correspond to the combinations of order and disorder parameters.

Let us first briefly discuss the analytic structure of the statistical mechanics model corresponding to $\langle \rho |\rho \rangle$. Recall that the Hamiltonian of the 2D AT model along the self-dual line depends only on one parameter, which we denote as $h$. Let's write the partition
function for a single copy of translationally invariant AT model as $\mathcal{Z}_{\text{AT}} = \sum_{t,z} \prod_e \omega_e(t, z; h)$ where explicitly, $\omega_e$ on edge $e$ of a square lattice is given by $\omega_e(t,z;h) = \left[\sqrt{2} + h + h (z_v z_{v'} + t_v t_{v'}) + (\sqrt{2} - h) z_v z_{v'} t_v t_{v'} \right]$. We find (see Appendix \ref{sec:app_derive_rhorho})

\bea
\langle \rho |\rho \rangle & = & \sum_{t,z,\bar{t},\bar{z}}\prod_e \left[\omega_e(t,z;h') \omega_e(\bar{t},\bar{z};h') \right.  \nonumber \\
& & \left. \hspace{0.9cm} + f(p)\, \omega_e(1/h',t,z) \omega_e(\bar{t},\bar{z};1/h') \right] \label{eq:nonfixedZ}
\eea 
where $h’ = 2h/(1+h^2)$ and $f(p) = 2p(1-p) h'^2/ (1-2p+2p^2)$. The edges $e$ belong to a square lattice. By expanding the product in Eq.\eqref{eq:nonfixedZ}, one can then write $\langle \rho |\rho \rangle$ as a sum of $2^{N_e}$ number of partition functions ($N_e$ = total number of edges on the underlying square lattice):

\be 
\langle \rho |\rho \rangle = \sum_{\{S\}} |f(p)|^{|\bar{S}|} \mathcal{Z}_S \bar{\mathcal{Z}}_S 
\ee 
where $\{S\}$ denotes the set of all distinct collection of edges, while $\mathcal{Z}_S$ and $\bar{\mathcal{Z}}_{S}$ are partition functions for the variables $\{z_{\mathbf{v}}, t_{\mathbf{v}} \}$ and $\{\bar{z}_{\mathbf{v}}, \bar{t}_{\mathbf{v}}\}$ respectively.  $\mathcal{Z}_{S} $ is explicitly given by  
\be 
\mathcal{Z}_{S}  = \sum_{t_{\mathbf{v}}, z_{\mathbf{v}}} \prod_{e \in S} \omega_e(z_{\mathbf{v}}, t_{\mathbf{v}}; h') \prod_{e \notin {S}} \omega_{e'}(z_{\mathbf{v}}, t_{\mathbf{v}}; 1/h') \label{eq:Z_S}
\ee 
and $\bar{\mathcal{Z}}_{S}$ has the identical form with $\{t_{\mathbf{v}}, z_{\mathbf{v}}\} \rightarrow \{\bar{t}_{\mathbf{v}}, \bar{z}_{\mathbf{v}}\}$. Therefore,  each $\mathcal{Z}_{S} $ corresponds to the partition function of a self-dual AT model with binary disorder (the translational invariance is restored due to the sum over $S$). More precisely, one can define a dual set of variables $\mu^z, \mu^t$ on the dual lattice and the self-duality implies that the partition function of the dual variables is identical to those of the original variables $z, t$ \cite{baxter2016exactly,wiseman1995critical}. This suggests  independent Kramers-Wannier duality for the fields $t_{\mathbf{v}}, z_{\mathbf{v}}$ and $\bar{t}_{\mathbf{v}}, \bar{z}_{\mathbf{v}}$ in the low energy theory, whose potential implications we will consider later.  Incidentally, from Eq.\eqref{eq:nonfixedZ}, one notices that when $h = 1$, $\langle \rho |\rho \rangle $ is completely \textit{independent} of $p$ up to a proportionality constant, and therefore, corresponds to four decoupled, critical Ising models for all $p$. 
We do not know an analytically tractable way to solve the partition function in Eqs.\eqref{eq:nonfixedZ},\eqref{eq:Z_S}. Therefore, we will instead use standard matrix product state (MPS) algorithms to carve out the global phase diagram in the $(h,p)$-plane for this problem.

Fig.\ref{Fig:nonfixed_phase}(a) shows the global phase diagram in the $(h,p)$ plane. Here BKT refers to the complex scalar CFT at the BKT point, while $\text{BKT}^2$ refers to two copies of the BKT theory i.e. two complex scalars at the BKT point. $\text{Ising}^4$ refers to four copies of the critical Ising theory. The position of the BKT point, i.e., $(h = 0, p = 0.5)$, is known from the exact mapping discussed in Sec.\ref{sec:stat_mech} while the positions of $\text{BKT}^2$ $(h = \sqrt{2}-1, p = 0)$ and $\text{Ising}^4$ $(h = 1, p = 0)$ are exactly known from Ref.\cite{zhu2019gapless}. We now discuss the details of how this phase diagram is obtained.

The basic idea behind the numerics is as follows. 
One can write $\langle \rho|\rho\rangle = \tr(T^{L_y})$ with $T$ the $2^{4 L_x} \times 2^{4 L_x}$ transfer matrix for the system of size $L_x \times L_y$. 
The dominant eigenvector of $T$, which we denote as $|\psi\rangle$, is equivalent to the ground state of the corresponding $(1+1)$-D Hamiltonian.
Assuming translational invariance, one can directly consider the thermodynamic limit $L_x, L_y \rightarrow \infty$ and then \textit{approximate} the ground state $|\psi\rangle$ using a translationally invariant MPS ansatz $|\psi_{\chi} \rangle $ with a finite bond dimension $\chi$.  
For a system with finite correlation length, the approximations of $|\psi_\chi \rangle$ as $|\psi\rangle$ converge rapidly as one increases $\chi$ \cite{eisert2010colloquim}. 
On the other hand, $|\psi_{\chi}\rangle$ can never faithfully represent the entanglement of true critical ground states, as its entanglement entropy $ S_{\text{vN}} $ of the bipartition on the infinite chain is bounded by $\log \chi$.
However, it turns out that the error of the approximation can be used to extract universal behaviors of the critical system \cite{pollmann2009theory}. In particular, the von-Neumann entropy $S_\text{vN}$, the bond dimension $\chi$, and the coorrelation length  $\xi$ of $|\psi_\chi\rangle$ are related to one another through $S_\text{vN} \approx c \kappa \log(\chi)/6 \approx c \log(\xi)/6$ with $c$ the central charge and $\kappa = 6/(\sqrt{12c} +c)$.
One can then consider a class of $|\psi_\chi\rangle$ with different $\chi$ to extract the central charge of the critical system, and such a method is called finite-entanglement scaling \cite{pollmann2009theory}.
In the following, we will first fix $\chi = 48$ to compute the correlation length of system in the $(h,p)$ plane [see Fig.\ref{Fig:nonfixed_phase}(c)].
The locations where $\xi$ starts to diverge signals the possibility of critical phase (note that for any finite $\chi$, the correlation length of $|\psi_\chi\rangle$ is always finite).
We will later use finite-entanglement scaling to provide more evidence of the existence of the critical phase [Fig.\ref{Fig:nonfixed_phase}(b)].

Fig.\ref{Fig:nonfixed_phase}(c) shows the inverse of correlation length $1/\xi$ as a function the error rate $p$ along the lines $h = 0.1$, $0.2$, and $0.3$.
When $p = 0$, we find that $1/\xi$ is some nonzero constant far from zero, indicating the system has finite correlation length.
This is consistent with the exact mapping that the system is gapped when $h<h_c = \sqrt{2} - 1 \approx 0.414$ \cite{zhu2019gapless}.
As one increases the error rate, $1/\xi$ for a given $h$  decreases and eventually becomes close to zero at a particular point $p_c(h)$, indicating the occurrence of a phase transition.
Besides, $1/\xi $ remains extremely small as one further increases the error rate, suggesting that the system remains critical above $p_c (h)$.
We also checked that for any $p$ at $h > h_c$, the system always has very large correlation length, suggesting that the system remains critical as one increases the error rate. Further, we confirmed that the gapped phase below the phase boundary is indeed a double topologically ordered state characterized by the uncondenesd and deconfined charge/flux anyon parameters. As mentioned above, for Fig.\ref{Fig:nonfixed_phase}(c) we used  $\chi = 48$. We also verified that similar results are obtained when one  further increases the bond dimension.

To confirm the presence of a critical phase outside the double  topologically ordered phase and to identify its universal features, we employ finite-entanglement scaling \cite{pollmann2009theory} to extract the central charge $c$.
We first benchmark the positions $(h,p) = (0.0,0.5)$ and $ (0.5,0.0)$, whose central charges are exactly known to be $c = 1$ and $c = 2$, respectively.
As shown in Fig.\ref{Fig:nonfixed_phase}(b), we find the numerically extracted central charges are $c \approx 0.87$ at $ (0.0,0.5)$ and $c \approx 1.86$ at $ (0.5,0.0)$, which are close to the exact values.
We then proceed to compute central charges at various points in the putative critical phase. We find that as long as the system is not too close to the BKT point $(h,p) = (0.0,0.5)$, the numerically extracted central charge is always around $c = 2$ [see Fig.\ref{Fig:nonfixed_phase}(b)].
Therefore, we expect the regime outside the double  topologically ordered phase is described by a central charge $c = 2$ theory except for the special point at $(h,p) = (0.0,0.5)$ which has central charge $c = 1$. Since the system is known to be described by two copies of the AT model at $p = 0$ for $ \sqrt{2} - 1 < h < 1$, which has a marginal deformation (the boson radius/Luttinger parameter), c-theorem \cite{zamolodchikov1986irreversibility} strongly indicates that the whole critical regime is described by two complex, free, massless scalars.

We also numerically considered the stability of the gapless phase against  perturbations that break the partial-transpose symmetry while retaining the EMD symmetry.
Such perturbations can be introduced by adding either the channel with Kraus operator a Pauli-$Y$ matrix or the combinations of phase-flip (Pauli-$Z$) and bit-flip (Pauli-$X$) channels with equal error rate. 
We applied these perturbations simultaneously with equal error rate of $p' = 0.1$ to the critical state at the location $(h,p) = (0.5,0.3)$, and find the numerically extracted central charge remains close to $2$.
This suggests that our critical phase is also stable under the perturbations that respect the EMD symmetry but break the partial-transpose symmetry.

\begin{figure}
	\centering
	\includegraphics[width=\linewidth]{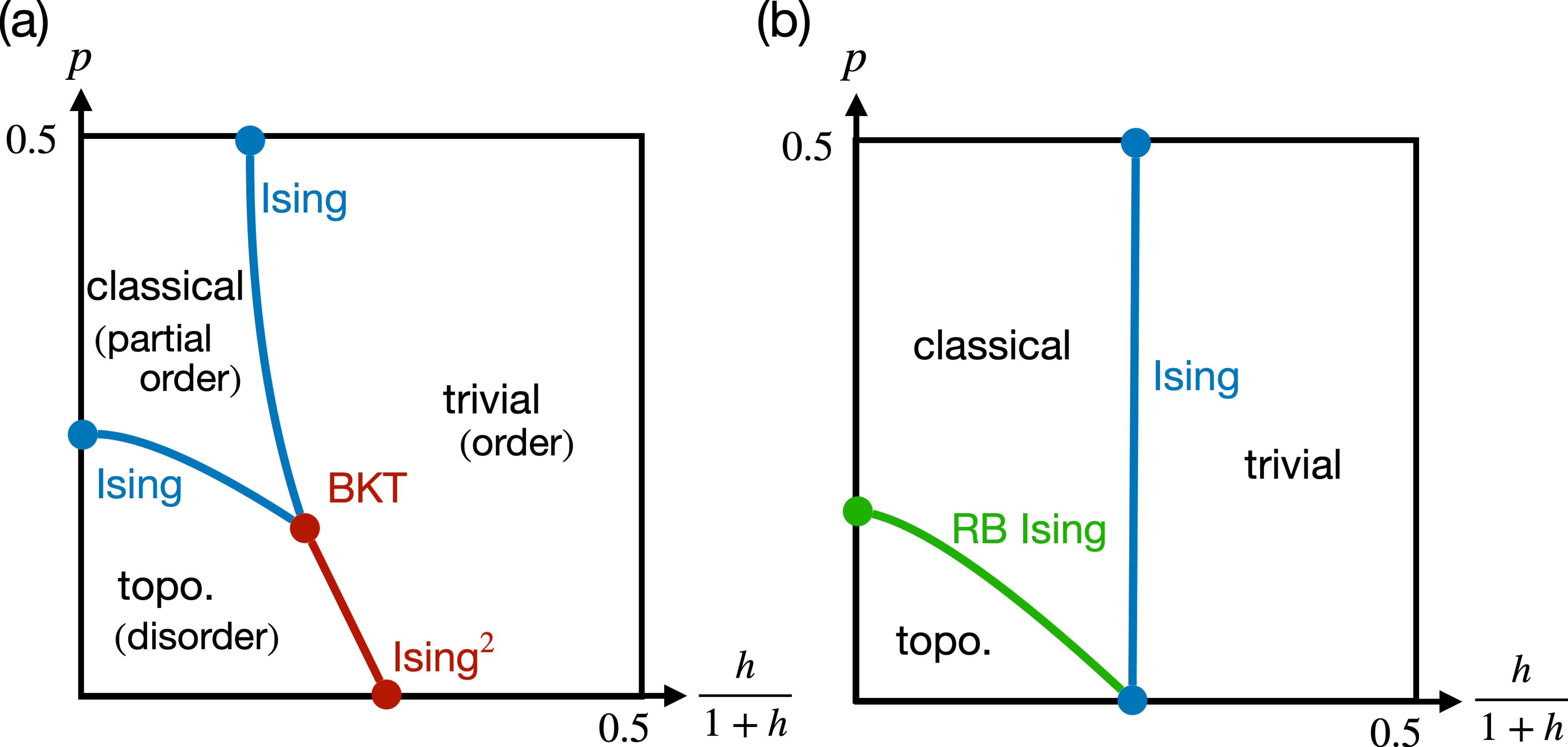}
	\caption{(a) The phase diagram for the perturbed toric code in Eq.\eqref{Eq:chamonstate} under phase-flip channel based on the quantities related to  $\log(\tr \rho^n )/(1-n)$ with $n = 2$. The central charge along the red line is $c = 1$, while that along the blue lines is $c =1/2$. (b) Conjectured phase diagram for the limit $n \rightarrow 1$. The location of the green circle along the $h = 0$  line is known from Refs. \cite{dennis2002,honecker2001universality}, the location of the blue circle along the $p = 0$ line is known from Ref.\cite{castelnovo2008quantum}, while the location of the other blue circle is determined in Appendix \ref{sec:chamon}.  
	}
	\label{Fig:chamon_phase}
\end{figure}

Given that we are able to explore new mixed-state phases and phase transitions by starting with a finite correlation length pure state, it is interesting to consider subjecting finite correlation length pure states to other quantum channels. To explore this direction, we also studied a differently perturbed toric code subjected to phase flip errors. In short, we consider the state 

\be 
|\Psi(h)\rangle = \prod_e (1+h Z_e) |\Psi_0\rangle \propto e^{K \sum_e Z_e}|\Psi_0\rangle  \label{Eq:chamonstate}
\ee 
subjected to the phase-flip channel, $\mathcal{E}[\cdot] = \prod_e \mathcal{E}_e [\cdot]$, where $\mathcal{E}_e [\cdot] = (1-p)\rho  + p Z_e \rho Z_e$. Interestingly, this case is analytically tractable and we provide the details in Appendix \ref{sec:chamon}.  The resulting phase diagram is schematically shown in Fig.\ref{Fig:chamon_phase}. One finds a rather rich phase diagram for the double state, with three phases, namely, a phase where both copies are separately topologically ordered (so that topological entanglement entropy (TEE) is $2 \log(2)$), a phase where $e \bar{e}$ is condensed (so that TEE is $\log(2)$), and a phase without any topological order.

Besides the unconventional phase transition out of the topologically ordered phase, perhaps the most striking feature in our phase diagram is the presence of a critical phase for a whole range of parameters (Fig.\ref{Fig:nonfixed_phase}). It is worth examining this phase and related aspects from the perspective of an effective field theory.
Let us start with the pure-state limit in the absence of decoherence (i.e. $p = 0$). The normalization  $\langle \rho|\rho \rangle = \langle \Psi(h) |  \Psi(h)\rangle \langle \bar{\Psi}(h)| \bar{\Psi}(h) \rangle $ can be written as a path integral whose Lagrangian corresponds to two decoupled sine-Gordon fields \cite{zhu2019gapless}:
\begin{equation}
	\label{Eq:sine_gordon}
	\mathcal{L}_{\text{sG}}  = \Big[ \frac{1}{2\pi K}((\partial_\tau \phi)^2  + (\partial_x \phi)^2 )+ \lambda \sin(4 \phi) \Big]+ \Big[ \phi \leftrightarrow \bar{\phi} \Big],
\end{equation}
where $K$ is the Luttinger parameter that depends on $h$, and $\phi (\bar{\phi})$ denotes the bosonic field in $\mathcal{H} (\bar{\mathcal{H}})$.
The connection between $\phi$ and the classical Ising spins $z_\mathbf{v}, t_\mathbf{v}$ that enter the partition function $\langle \rho |\rho \rangle$ is through the Majorana spinors $\gamma_R + i \eta_R \sim e^{i (\phi + \theta) }  $, $\gamma_L + i \eta_L \sim e^{- i (\phi - \theta) } $, where $\theta$ is the field dual to $\phi$. Here $\gamma= (\gamma_{L}, \gamma_{R})$ [$\eta= (\eta_{L}, \eta_{R})$] correspond to the Ising spins $z_\mathbf{v}$($t_\mathbf{v}$) in the standard way  \cite{schultz1964two, jordan1993paulische}: the transfer matrix for the critical Ising model corresponds to a (1+1)-D transverse field Ising model at criticality, which can be fermionized via the Jordan-Wigner transformation. Similar relations hold for $\bar{\phi}$ and  $\bar{z}_\mathbf{v}, \bar{t}_\mathbf{v}$.  EMD symmetry corresponds to the lattice translation for  Majorana fermions, which in the continuum limit acts as  $(\gamma_L, \eta_L, \bar{\gamma}_L, \bar{\eta}_L) \rightarrow -(\gamma_L, \eta_L, \bar{\gamma}_L, \bar{\eta}_L)$ while the right-moving Majorana fermions remain invariant.
The gapless regime $h \in [\sqrt{2} - 1, 1]$ corresponds to $K \in [1/2, 1]$ such that the scaling dimension of $\sin(4 \phi) = 4K < 2$, and thus this term is irrelevant.
In particular, the free-fermion limit $K = 1$ corresponds to $h = 1$, as $\mathcal{N} (h )= \prod_e (I + h \sigma_e)$ becomes a projector that decouples $\gamma (\bar{\gamma})$ and $\eta (\bar{\eta})$.
On the other hand, the double topologically ordered phase corresponds to $K < 1/2$ (i.e. $ 0 \leq h < \sqrt{2}-1$) and $\lambda > 0$ such that $\sin(2 \phi) \sim  i (\gamma_L \eta_R - \eta_L \gamma_R)$ and $\sim  i (\bar{\gamma}_L \bar{\eta}_R - \bar{\eta}_L \bar{\gamma}_R) $ are pinned to $\pm 1$.
This can be seen by recalling that the double topological phase corresponds to the double partial order phase in the statistical model (see Sec.\ref{sec:coherent}), and the partial order phase can be detected by the operator that is even under the simultaneous spin-flip action $\gamma, \eta \rightarrow -\gamma, -\eta$ but odd under the single spin-flip operation $\gamma \rightarrow -\gamma$ or $\eta \rightarrow -\eta$.

Next, let's discuss the physics of the critical phase (Fig.\ref{Fig:nonfixed_phase}). As discussed above, numerically we find that the central charge of the statistical mechanics model for the second Renyi entropy $\tr \rho^2 = \langle \rho |\rho \rangle$ throughout the critical phase equals  two within our numerical accuracy. Our expectation is that the low-energy theory in this phase is again described by Eq.\eqref{Eq:sine_gordon} where the Luttinger parameter will generically be a function of the decoherence rate $p$ and the parameter $h$ that defines the pure state (Eq.\eqref{Eq:selfdual_nonfixed}). Here we encounter a puzzle whose resolution we don't fully understand. The decohered density matrix only respects the EMD symmetry that acts on both copies simultaneously, i.e., $\mathbb{U}_D^{\dagger} \rho \mathbb{U}_D = \rho$ where $\mathbb{U}_D =  U_D \bar{U}_D$ (see Eq.\eqref{eq:U_D} for the definition of $U_D$). Therefore, one might expect that the following operator, which is invariant under $\mathbb{U}_D$ but odd under $U_D$ and $\bar{U}_D$, is allowed:

\begin{equation}
	\label{Eq:Lint2}
	\Delta \mathcal{L} = \lambda' (\gamma_{L} \gamma_R + \eta_L \eta_R) (\bar{\gamma}_L \bar{\gamma}_R +\bar{\eta}_L \bar{\eta}_R ).
\end{equation}

After bosonizing Eq.\eqref{Eq:Lint2}, $\Delta \mathcal{L}  = \lambda' \cos(2 \phi) \cos(2 \bar{\phi})$, which has scaling dimension $\text{dim}[\cos(2 \phi) \cos(2 \bar{\phi})] = 2 K$. 
When $p = 0$ and $h \in [\sqrt{2}-1, 1)$, one finds that $K \in [1/2, 1)$ and thus $\Delta \mathcal{L} $ is relevant. This will lead to spontaneous symmetry breaking resulting in $\cos(2\phi) = \cos(2 \bar{\phi}) = \pm 1$. This is at variance with our numerical observation that the critical phase is in fact stable everywhere in the phase diagram outside the topological phase. First, let's briefly consider the consequences for the low-energy physics if $\Delta \mathcal{L} $ were indeed present in the effective field theory. One expects that $\Delta \mathcal{L} $ being relevant will pin the system to a state that breaks the EMD symmetry spontaneously, and satisfies $\cos(2\phi) = \cos(2 \bar{\phi}) = \pm 1$ where $+1$ and $-1$ correspond to the condensation of flux and charge respectively. This implies that the true ground state will be a superposition of states corresponding to flux and charge condensates. Such a state will violate the standard intuition that anyons with non-trivial mutual statistics can't condense at the same time. Relatedly, the ground state of the transfer matrix of the corresponding statistical mechanics model will have simultaneous existence of order and disorder parameters, which is rather unusual but still allowed if certain conditions are satisfied \cite{o2018lattice,levin2020constraints}. We did not find evidence for such a state in our phase diagram (Fig.\ref{Fig:nonfixed_phase}). However, it is indeed possible that in the thermodynamic limit, the ground state is gapped with a very large correlation length so that it appears critical for the maximum bond dimension we used.

Given the stability of the critical phase observed in our numerics, we  suspect that $\Delta \mathcal{L}  $ is in fact not allowed at low-energies. The absence of $\Delta \mathcal{L}$ in the low energy theory is suggested by Eqs.\eqref{eq:nonfixedZ},\eqref{eq:Z_S}, where we find $\langle \rho|\rho \rangle$ can be expressed as sum of partition functions that, despite being non-translationally invariant, are self-dual under separate Kramers-Wannier duality for the fields $\{t_{\mathbf{v}}, z_{\mathbf{v}}\}$ and $\{\bar{t}_{\mathbf{v}}, \bar{z}_{\mathbf{v}}\}$. This suggests that one may write the continuum version of the partition function as

\bea
\label{Eq:sine_gordon_disorder}
\mathcal{Z} & = & \int D\phi \int DK\,D\lambda\, P(K,\lambda) e^{- \int dx d\tau \mathcal{L}} \nonumber \\
\mathcal{L} & = & \Big[ \frac{1}{2\pi K(x,\tau)}\left((\partial_\tau \phi(x,\tau))^2  + (\partial_x \phi(x,\tau))^2 + \Big[ \phi \leftrightarrow \bar{\phi} \Big]\right) \Big. \nonumber \\
& & +\Big. \lambda(x,\tau) \left(\sin(4 \phi(x,\tau))+  \Big[ \phi \leftrightarrow \bar{\phi} \Big] \right)  \Big] 
\eea
where $P(K,\lambda)$ is the probability distribution for the couplings $K$ and $\lambda$ and is the continuum analog of the  factor $|f(p)|^{|\bar{S}|} $ in the lattice partition function (see Eq.\eqref{eq:nonfixedZ}), and integrating over $K,\lambda$ will generate interactions between the fields $\phi$ and $\bar{\phi}$. The partition function in Eq.\eqref{Eq:sine_gordon_disorder} clearly possesses a `doubled Kramers-Wannier' symmetry corresponding to independently sending $\phi \rightarrow \phi + \pi/2$, or $\bar{\phi} \rightarrow \bar{\phi} + \pi/2$. We leave further investigation of this interesting direction to the future.

Assuming that $\Delta \mathcal{L} $ is not allowed in the effective action, one can then understand the phase transition between the topologically ordered phase and the critical phase in terms of the most relevant operator allowed by symmetries, namely, $\sin(4 \phi)$. Approaching this transition from the critical phase, as decoherence decreases, one expects that the Luttinger parameter $K$ will decrease and eventually approach $K_c = 1/2$ so that $\sin(4 \phi)$ becomes relevant.  This is because as discussed in Sec.\ref{sec:stat_mech}, the effect of decoherence will not only introduce the coupling between the fields in $\mathcal{H}$ and $\bar{\mathcal{H}}$ (corresponding to a new term that couples $\phi$ and $\bar{\phi}$) but also \textit{decrease} the coupling between the two copies. This implies decreasing the error rate is expected to decrease the Luttinger parameter $K$. This qualitatively explains the phase boundary between the topological phase and the critical phase  in our phase diagram (Fig.\ref{Fig:nonfixed_phase}).%

\section{Summary and discussion} \label{sec:discuss}
In this work we obtained  constraints on anyon condensation in mixed states of toric code by imposing weak EMD and partial-transpose symmetries. Throughout, we focussed on the double state corresponding to the mixed state. Our conclusion is that in the presence of EMD and partial-transpose symmetry, conventional anyon condensation mechanism, where string operators corresponding to anyons take non-zero expectation value, is ruled out. We derived this result using a combination of symmetry arguments and the triangle inequality. To illustrate the implications of this constraint, we provided an example where a phase transition out of the double topologically ordered  state is necessarily unconventional in the sense that it cannot occur via conventional anyon condensation. In particular, we studied a finite-correlation version of toric code (Eq.\eqref{Eq:selfdual_nonfixed}), and subjected it to a channel that preserves  EMD and partial-transpose symmetry. Above a critical strength, the topological phase is destroyed via an unconventional transition, and we find numerical evidence for a critical phase where anyons are only power-law condensed, i.e., the string operators that characterize anyon condensation have a power-law scaling. Our numerical results and analytical arguments are consistent with the conclusion that these correlations are described by a CFT corresponding to two free, massless, complex scalars. The phase diagram is schematically shown in Fig.\ref{Fig:fig1} with more detailed aspects shown in Fig.\ref{Fig:nonfixed_phase}. We numerically find that the critical phase is stable against breaking the partial-transpose symmetry and only requires the EMD symmetry. For the special case corresponding to the zero correlation length toric code subjected to the same channel, we find that the double  topologically ordered state is stable up to the maximal decoherence strength, and right at the maximal decoherence strength, the topological ordered phase is destroyed and the system possesses critical correlations corresponding to a single free, massless, complex scalar (Figs.\ref{Fig:coherent_phase},\ref{Fig:nonfixed_phase}). For this case, the  statistical mechanics model corresponding to the normalization of the double state  is mapped to a single copy of the Ashkin-Teller model along the self-dual line in the critical regime. For the more general case of the aforementioned, decohered perturbed toric code, the statistical mechanics model can be expressed in terms of two coupled AT models along the self-dual line.

In this work we only focussed on the double state subjected to $X+Z$ Kraus operators. It will be very interesting to study the intrinsic properties of the mixed state corresponding to a single copy, such as aspects related to error-correction, mixed-state entanglement and separability \cite{dennis2002,wang2003confinement, lee2023quantum, fan2023diagnostics,bao2023mixed,chen2023separability,lee2024exact}. For example, in the case of toric code decohered by phase-flip errors, the double state undergoes a transition in the translationally invariant Ising universality, while the universal aspects of error correction and entanglement for the single copy are related to the random bond Ising model along the Nishimori line. For our problem, the universality for the double state transition is a free, complex scalar, and it is interesting to speculate if the single copy transition belongs to a random version of this problem along the Nishimori line. We leave this problem for future investigation.

Another interesting direction is to subject the ground state of $H = H_{\text{TC}} - h \sum (X_i + Z_i)$ to the channel with Kraus operators $X_i + Z_i$. Would the corresponding double state exhibit a phase diagram similar to that in Fig.\ref{Fig:nonfixed_phase}? It will also be interesting to consider self-dual decoherence in models where the EMD symmetry is onsite and does not involve translations, such as those studied in \cite{cheng2017exactly}. Such models will allow one to enforce strong EMD symmetry and may have a richer phase diagram. Another potential direction is to generalize the arguments in Sec.\ref{sec:condition} to more general topologically ordered states with EMD symmetry such as $S_3$ gauge theory \cite{beigi2011quantum}.

\textit{Note added:} While this manuscript was being finalized, Ref.\cite{eckstein2024robust} appeared which has overlap with our Sec.III where we study the fixed-point toric code decohered by $X + Z$ Kraus operators and its connection with the self-dual Ashkin-Teller model. Although our  focus and observables  studied are different from Ref.\cite{eckstein2024robust}, the qualitative features do seem to match where they overlap.

\begin{acknowledgments}
We thank John McGreevy for helpful comments on the manuscript. TG is supported by the National Science Foundation under Grant No. DMR-1752417.
\end{acknowledgments}

%


\onecolumngrid
\appendix

\section{Statistical mechanical mapping for toric code under local decoherence}

The goal of this Appendix is to discuss the statistical mechanical mapping for the toric code under local decoherence.
We will show in Appendix \ref{sec:app_anyon_mapping} that the anyon overlap parameters discussed in Sec.\ref{sec:condition} can always be mapped to some combinations of ordered and disordered parameters in the corresponding classical model.
In Appendix \ref{sec:app_coherent}, we will focus on the mapping for the coherent channel discussed in the main text.
In particular, we will map the second moment of the density matrix (equivalent to the normalization of the double state) to the partition function of the isotropic Ashkin-Teller model.

\subsection{General mapping for anyon overlap parameters}
\label{sec:app_anyon_mapping}

We now show that the anyon overlap parameters 
$\langle {\alpha \bar{\beta}} | {\gamma \bar{\delta}} \rangle \equiv \langle \rho_{\alpha \bar{\beta}} | \rho_{\gamma \bar{\delta}} \rangle / \langle \rho | \rho\rangle$ can be mapped to various combinations of order and disorder parameters of the spin variables $(z,  \bar{z})$  and  $(t,  \bar{t})$.
To begin with, we first use Eq.\eqref{Eq:toric_wavefunction} to write down the initial double state:
\begin{equation}
	\begin{aligned}
		|\rho_0\rangle = |\Psi_0\rangle |\bar{\Psi}_0\rangle 
		=  \sum_{x_\mathbf{e}, \bar{x}_\mathbf{e}} [\sum_{z_v, \bar{z}_v} \prod_e (1+x_e z_v z_{v'}) (1+\bar{x}_e \bar{z}_v \bar{z}_{v'})] |x_\mathbf{e}, \bar{x}_\mathbf{e} \rangle.
	\end{aligned}
\end{equation}
One can also easily write down the initial anyon state $|\rho_{\alpha \beta, 0} \rangle = w_{\alpha} \bar{w}_{\beta}|\rho_0\rangle$.
Take $|\rho_{e \bar{e},0}\rangle$ as an example.
Since a pair of $e \bar{e}$ anyon state located at sites $i$ and $j$  can be created by any string operator $w_e(l) \bar{w}_e(l) = \prod_{e \in l} X_e \bar{X}_e$ with the end points located at sites $i$ and $j$, one finds
\begin{equation}
	\begin{aligned}
		\label{Eq:charge_anyon_double}
		|\rho_{e\bar{e},0} \rangle &  
		= \prod_{e \in l} X_e \bar{X}_e |\rho_0\rangle 
		= 
		\sum_{x_\mathbf{e}, \bar{x}_\mathbf{e}} \sum_{z_\mathbf{v}, \bar{z}_\mathbf{v}} \prod_{e \in l} (x_e+ z_v z_{v'}) (\bar{x}_e+ \bar{z}_v \bar{z}_{v'})
		\prod_{e \notin l} (1+x_e z_v z_{v'}) (1+\bar{x}_e \bar{z}_v \bar{z}_{v'}) |x_e, \bar{x}_e\rangle 
		\\
		& = 
		\sum_{x_\mathbf{e}, \bar{x}_\mathbf{e}e} [\sum_{z_\mathbf{v}, \bar{z}_\mathbf{v}} \prod_e z_i z_j \bar{z}_i \bar{z}_j(1+x_e z_v z_{v'}) (1+\bar{x}_e \bar{z}_v \bar{z}_{v'})] |x_e, \bar{x}_e\rangle \\
		& = \sum_{z_\mathbf{v}, \bar{z}_\mathbf{v}} z_i z_j \bar{z}_i \bar{z}_j \sum_{x_\mathbf{e}, \bar{x}_\mathbf{e}} [ \prod_e (1+x_e z_v z_{v'}) (1+\bar{x}_e \bar{z}_v \bar{z}_{v'})] |x_e, \bar{x}_e\rangle.
	\end{aligned}
\end{equation}
In the second line, we use  $(x_e+ z_v z_{v'}) (\bar{x}_e+ \bar{z}_v \bar{z}_{v'}) = z_v z_{v'} \bar{z}_v \bar{z}_{v'} (1+x_e z_v z_{v'}) (1+\bar{x}_e \bar{z}_v \bar{z}_{v'}) $ and the property that all the spins $z_v$ with $v \notin i,j$ appear even number of times.
In the third line, we simply change the order of the summations $ \sum_{x_\mathbf{e}, \bar{x}_\mathbf{e}}$ and $ \sum_{z_\mathbf{v}, \bar{z}_\mathbf{v}} $ so that it becomes transparent that the effect of the operator $\prod_{e \in l} X_e \bar{X}_e$ on $|\rho_0\rangle$ is transformed  to $z_i z_j \bar{z}_i \bar{z}_j$, the two-point function of the matter field.
Now, let's consider the $e \bar{e}$ condensation parameters ${\langle \rho_0|\mathcal{E}^\dagger \mathcal{E} |\rho_{e \bar{e},0} \rangle} /{ \langle \rho_0|\mathcal{E}^\dagger \mathcal{E}|\rho_0\rangle}$. 
As already shown in the main text,
$ \langle \rho_0 | \mathcal{E}^\dagger \mathcal{E}|\rho_0\rangle  = \sum_{z_\mathbf{v}, \bar{z}_\mathbf{v}, t_\mathbf{v},  \bar{t}_\mathbf{v}} e^{-H(z_\mathbf{v}, \bar{z}_\mathbf{v}, t_\mathbf{v}, \bar{t}_\mathbf{v})}$, where
\begin{equation}
	\begin{aligned}
		e^{-H(z_\mathbf{v}, \bar{z}_\mathbf{v}, t_\mathbf{v}, \bar{t}_\mathbf{v})} &  = \sum_{\substack{x_\mathbf{e}, \bar{x}_\mathbf{e}, \\
				x'_\mathbf{e}, \bar{x}'_\mathbf{e} }}
		\Big[ \langle x'_\mathbf{e}, \bar{x}'_\mathbf{e} |{\mathcal{E}}^\dagger {\mathcal{E}}|x_\mathbf{e}, \bar{x}_\mathbf{e}\rangle   \prod_e (1 + x_e z_v z_{v'}) (1 + \bar{x}_e \bar{z}_v \bar{z}_{v'}) (1 + x'_e t_v t_{v'}) (1 + \bar{x}'_e \bar{t}_v \bar{t}_{v'}) 
		\Big].
	\end{aligned}
\end{equation}
Since $z_i z_j \bar{z}_i \bar{z}_j$ is outside the summation of $\sum_{x_\mathbf{e}, \bar{x}_\mathbf{e}}$, one finds $ \langle \rho_0 | \mathcal{E}^\dagger \mathcal{E}|\rho_{e\bar{e},0}\rangle  = \sum_{z_\mathbf{v}, \bar{z}_\mathbf{v}, t_\mathbf{v},  \bar{t}_\mathbf{v}} z_i z_j \bar{z}_i \bar{z} e^{-H(z_\mathbf{v}, \bar{z}_\mathbf{v}, t_\mathbf{v}, \bar{t}_\mathbf{v})}$, and thus $\langle I \bar{I} |e \bar{e}\rangle$ corresponds to the two-point correlation function of $z \bar{z}$ no matter what the channel $\mathcal{E}$ is.
A straightforward generalization to other anyon states related to charges then show that we can identify $(e, \bar{e})$ in $|\Psi_0\rangle |\bar{\Psi}_0\rangle$ as $(z,  \bar{z})$ and $(e, \bar{e})$ in $\langle \Psi_0| \langle \bar{\Psi}_0|$  as $(t,  \bar{t})$.

On the other hand, a pair of $m \bar{m}$ ayons located at sites $\tilde{i}$ and $\tilde{j}$ on the dual lattice can be created by $w_m(\tilde{l}) \bar{w}_m (\tilde{l}) = \prod_{e \in \tilde{l}} Z_e \bar{Z}_e$, where $\tilde{l}$ can be any string with end points located at $\tilde{i}$ and $\tilde{j}$.
It follows that
\begin{equation}
	\label{Eq:flux_double}
	\begin{aligned}
		|\rho_{m \bar{m}, 0} \rangle & = \prod_{e \in \tilde{l}} Z_e \bar{Z}_e |\rho_0\rangle=  \sum_{x_\mathbf{e}, \bar{x}_\mathbf{e}} [\sum_{z_\mathbf{v}, \bar{z}_\mathbf{v}}\prod_e (1+x_{e} z_v z_{v'}) (1+\bar{x}_{e} \bar{z}_v \bar{z}_{v'})] |\eta_{\mathbf{e}} x_\mathbf{e}, \eta_{\mathbf{e}}  \bar{x}_\mathbf{e}\rangle \\
		& =  \sum_{x_\mathbf{e}, \bar{x}_\mathbf{e}} [\sum_{z_\mathbf{v}, \bar{z}_\mathbf{v}}\prod_e (1+\eta_{e} x_{e} z_v z_{v'}) (1+\eta_{e} \bar{x}_{e} \bar{z}_v \bar{z}_{v'})] |x_\mathbf{e}, \bar{x}_\mathbf{e}\rangle,
	\end{aligned}
\end{equation}
where $\eta_{e} = 1 (-1)$ if $e \notin \tilde{l} (\in \tilde{l})$.
In the second line, we relabel the variables $x_\mathbf{e} (\bar{x}_\mathbf{e})$ as $ \eta_\mathbf{e} x_\mathbf{e}  (\eta_\mathbf{e} \bar{x}_\mathbf{e})$.
Therefore, the effect of $\prod_{e \in \tilde{l}} Z_e \bar{Z}_e$ on $|\rho_0\rangle$ corresponds to changing the sign of the Ising interaction for $z_\mathbf{v}$ and $\bar{z}_\mathbf{v}$. 
It is then obvious that $\langle \rho_0 |\mathcal{E}^\dagger \mathcal{E} |\rho_{m \bar{m},0}\rangle = \sum_{z_\mathbf{v}, \bar{z}_\mathbf{v}, t_\mathbf{v}, \bar{t}_\mathbf{v} } e^{- H_{ z_{\tilde{l}}\bar{z}_{\tilde{l}}} (z_\mathbf{v}, \bar{z}_\mathbf{v}, t_\mathbf{v}, \bar{t}_\mathbf{v} )} $,
where
\begin{equation}
	\begin{aligned}
		e^{- H_{ z_{\tilde{l}}\bar{z}_{\tilde{l}}}(z_\mathbf{v}, \bar{z}_\mathbf{v}, t_\mathbf{v}, \bar{t}_\mathbf{v})  }  = \sum_{\substack{x_\mathbf{e}, \bar{x}_\mathbf{e}, \\
				x'_\mathbf{e}, \bar{x}'_\mathbf{e} }}
		\Big[ \langle x'_\mathbf{e}, \bar{x}'_\mathbf{e} |{\mathcal{E}}^\dagger {\mathcal{E}}|x_\mathbf{e}, \bar{x}_\mathbf{e}\rangle \times 
		\prod_e  (1 + \eta_e x_e z_v z_{v'}) (1 + \eta_e \bar{x}_e \bar{z}_v \bar{z}_{v'}) (1 + x'_e t_v t_{v'}) (1 + \bar{x}'_e \bar{t}_v \bar{t}_{v'}) 
		\Big].
	\end{aligned}
\end{equation}
Therefore, $\langle \rho_0 |\mathcal{E}^\dagger \mathcal{E} |\rho_{m \bar{m},0}\rangle / \langle \rho_0 |\mathcal{E}^\dagger \mathcal{E} |\rho_{0}\rangle $ is nothing but the disorder parameter $\langle \mu^{z}_{\tilde{i}} \mu^{\bar{z}}_{\tilde{i}} \mu^{z}_{\tilde{j}} \mu^{\bar{z}}_{\tilde{j}} \rangle$.
A straightforward generalization to other anyon states related to flux then show that we can identify $(m, \bar{m})$ in $|\Psi_0\rangle |\bar{\Psi}_0\rangle$  as $(\mu^z,  \mu^{\bar{z}})$ and $(m, \bar{m})$ in $\langle \Psi_0| \langle \bar{\Psi}_0|$  as $(\mu^t,  \mu^{\bar{t}})$.

\subsection{Coherent channel}
\label{sec:app_coherent}
We now focus on the specific coherent error channel  $\mathcal{E}[\cdot] = \prod_e \mathcal{E}_e[\cdot]$, where $\mathcal{E}_e[\cdot] = (1-p)(\cdot) + p\sigma_e (\cdot) \sigma_e,\ \sigma_e = \cos(\theta)Z_e + \sin(\theta)X_e$.
After C-J map, one finds $\mathcal{E} = \mathcal{E}(p, \theta) = \prod_e [(1-p)+p\sigma_e(\theta) \bar{\sigma}_e(\theta)]$.
To simplify the calculation of the normalization $\langle \rho | \rho \rangle = \langle \rho_0 | \mathcal{E}^\dagger  \mathcal{E} |\rho_0\rangle$, we note that that since $ \mathcal{E}(p, \theta)^\dagger  \mathcal{E}(p, \theta) = \mathcal{E}(p', \theta)$ with $p' = 2p(1-p)$, one finds $\langle \rho_0 | \rho(p',\theta) \rangle = \langle \rho (p, \theta)| \rho(p,\ \theta) \rangle$.
Therefore, instead of directly evaluating $\langle \rho (p, \theta)| \rho(p,\ \theta) \rangle$, we will compute $\langle \rho_0 | \rho(p',\theta) \rangle$, which is relatively simpler to work with.
Now, a direct substitution of $\mathcal{E}(p', \theta)$ gives
\begin{equation}
	\langle \rho_0 | \rho(p',\theta) \rangle  = \langle \rho_0 |  \prod_e [[1-2p(1-p)]I + 2p(1-p)\sigma_e \bar{\sigma}_e]| \rho_0 \rangle  
	\propto
	\langle \rho_0 |  \prod_e (I +\lambda \sigma_e \bar{\sigma}_e)| \rho_0 \rangle  ,
\end{equation}
where $\lambda = 2p(1-p)/ (1-2p+2p^2)$ and  
\begin{equation}
	\sigma_e(\theta)\bar{\sigma}_e(\theta)=\sin^2(\theta) X_e \bar{X}_e+ \sin(\theta) \cos(\theta) (X_e\bar{Z}_e + \bar{Z}_e X_e) + \cos^2(\theta)Z_e \bar{Z}_e.
\end{equation}
Using the property that   $\langle \Psi_0 | \prod_{e \in l}X_e |\Psi_0\rangle = 0$ ($\langle \Psi_0 | \prod_{e \in \tilde{l}} Z_e |\Psi_0\rangle = 0$) unless  $l $ ($\tilde{l}$) forms closed loops in the original (dual) lattice, one finds that when expanding $\prod_e (I + \lambda \sigma_e \bar{\sigma}_e)$, any terms that include $(X_e\bar{Z}_e + \bar{Z}_e X_e)$ take zero expectation value with $|\rho_0\rangle = |\Psi_0\rangle |\Psi_0\rangle$.
Therefore, we only need to consider $|\rho'(\lambda,\theta)\rangle = \prod_e [1+\lambda(\sin^2(\theta) X_e \bar{X}_e + \cos^2(\theta)Z_e \bar{Z}_e)]|\rho_0\rangle$, and a straightforward algebra gives
$|\rho'(\lambda, \theta)\rangle  =  \sum_{x_e, \bar{x}_e} [\sum_{z_v, \bar{z}_v}  \prod_e \nu_e (z_v, \bar{z}_v)] |x_e, \bar{x}_e \rangle$, where
\begin{equation}
	\begin{aligned}
		\nu_e  =  (1+\lambda \sin^2(\theta) x_e\bar{x}_e )(1+x_e z_v z_{v'})(1+\bar{x}_e \bar{z}_v \bar{z}_{v'}) 
		+ \lambda \cos^2(\theta) (1-x_e z_v z_{v'})(1-\bar{x}_e \bar{z}_v \bar{z}_{v'}) . 
	\end{aligned}
\end{equation} 
Now, using $\langle \rho_0 | x_e, \bar{x}_e\rangle =  [\sum_{t_v, \bar{t}_v} \prod_e (1+x_e t_v t_{v'}) (1+\bar{x}_e \bar{t}_v \bar{t}_{v'})]$, one finds $	\langle \rho_0|\rho(\lambda, \theta)\rangle = \sum_{x_e, \bar{x}_e} \langle \rho_0 | x_e, \bar{x}_e\rangle  \langle   x_e, \bar{x}_e| \rho'(\lambda,\theta)\rangle \propto \sum_{z_\mathbf{v}, \bar{z}_\mathbf{v}, t_\mathbf{v}, \bar{t}_\mathbf{v}} \prod_e \omega_e(z_\mathbf{v}, t_\mathbf{v})$, where 
\begin{equation}
	\label{Eq:omega_e_XZ}
	\begin{aligned}
		\omega_e = & 1+ z_v z_{v'} \bar{z}_v \bar{z}_{v'} t_v t_{v'}\bar{t}_v \bar{t}_{v'}  + \frac{1- \lambda \cos^2(\theta)}{1+ \lambda \cos^2(\theta) } (t_v t_{v'}z_v z_{v'} +\bar{t}_v \bar{t}_{v'}\bar{z}_v \bar{z}_{v'}  )
		\\
		& 
		+ \frac{\lambda \sin^2(\theta)}{1+\lambda \cos^2(\theta)} (t_v t_{v'}\bar{t}_v \bar{t}_{v'} + z_v z_{v'}\bar{z}_v \bar{z}_{v'} +t_v t_{v'} \bar{z}_v \bar{z}_{v'} + \bar{t}_v \bar{t}_{v'} z_v z_{v'}) 
	\end{aligned}
\end{equation}
One can simplify Eq.\eqref{Eq:omega_e_XZ} by noting that $\omega_e z_v z_{v'} \bar{z}_v \bar{z}_{v'} t_v t_{v'}\bar{t}_v \bar{t}_{v'}  =\omega_e$ so that $\langle z_v z_{v'} \bar{z}_v \bar{z}_{v'} t_v t_{v'}\bar{t}_v \bar{t}_{v'} \rangle = 1, \forall e$.
Therefore, we can replace $\bar{t}_v \bar{t}_{v'} $ as $z_v z_{v'} \bar{z}_v \bar{z}_{v'} t_v t_{v'}$, and  Eq.\eqref{Eq:omega_e_XZ} becomes
\begin{equation}
	\omega_e \propto   1+ \frac{\lambda \sin^2(\theta)}{1+\lambda \cos^2(\theta)} (z_v z_{v'}\bar{z}_v \bar{z}_{v'} +t_v t_{v'} \bar{z}_v \bar{z}_{v'} )   + \frac{1- \lambda \cos^2(\theta)}{1+ \lambda \cos^2(\theta) } t_v t_{v'}z_v z_{v'}.
\end{equation}
Defining the variables $s_v = z_v \bar{z}_v,\ \tau_v = t_v \bar{z}_v$, $\omega_e$ can be further simplified to 
\begin{equation}
	\label{Eq:omega_ZX_simple}
	\begin{aligned}
		\omega_e & \propto   1+ \frac{\lambda \sin^2(\theta)}{1+\lambda \cos^2(\theta)} (s_v s_{v'} +\tau_v \tau_{v'} )  + \frac{1- \lambda \cos^2(\theta)}{1+ \lambda \cos^2(\theta) } s_v s_{v'}\tau_v \tau_{v'} \\
	\end{aligned}
\end{equation}
By reexponentiating Eq.\eqref{Eq:omega_ZX_simple},  one finds
\begin{equation}
	\omega_e(z_\mathbf{v}, t_\mathbf{v}) \propto e^{K (s_v s_{v'} +\tau_v \tau_{v'} ) + K_4( s_v s_{v'}\tau_v \tau_{v'})}.
\end{equation} 
which is nothing but the Ashkin-Teller model with the coeffecient  $K$ and $K_4$ related to $h'$ and $\theta$ through
\begin{equation}
	\begin{aligned}
		\frac{\tanh(K) [1+\tanh(K_4)] }{1+\tanh^2(K) \tanh(K_4)}  =   \frac{\lambda \sin^2(\theta)}{1+\lambda \cos^2(\theta)} , \ 
		\frac{\tanh^2(K) +\tanh(K_4)}{1+\tanh^2(K) \tanh(K_4)}  =  \frac{1- \lambda \cos^2(\theta)}{1+ \lambda \cos^2(\theta) } .
	\end{aligned}
\end{equation}
In particular, when $\theta = \pi/4$, one finds 
\begin{equation}
	\begin{aligned}
		\tanh(K)  = \frac{2- \sqrt{4-{\lambda}^2}}{\lambda},\
		\tanh(K_4)  = \frac{2- h\sqrt{4-{\lambda}^2}}{2-{\lambda}^2}.
	\end{aligned}
\end{equation}

\section{Derivation of Eq.\eqref{eq:nonfixedZ}}
\label{sec:app_derive_rhorho}
We here derive the compact form of the normalization of the decohered double state $|\rho(h,p) \rangle = \mathcal{E}(p) \mathcal{N}(h)  \bar{\mathcal{N}}(h) |\Psi_0, \bar{\Psi}_0 \rangle$ in Eq.\eqref{eq:nonfixedZ}, where $\mathcal{E}(p) = \prod_e [(1-p) + p \sigma_e \bar{\sigma}_e]$,  $\mathcal{N}(p) = \prod_e (1 + h \sigma_e) $, and $\sigma_e = (X_e + Z_e)/\sqrt{2}$.
We first make use of the property that $[\mathcal{E}, \mathcal{N}] = 0$, $\mathcal{E}(p)^\dagger \mathcal{E}(p)= \mathcal{E}(p')$ with $p' = 2p(1-p) $, and $\mathcal{N}(h)^\dagger \mathcal{N}(h) \propto \mathcal{N}(h')$ with $h' = 2h/(1+h^2)$ to write 
$\mathcal{N}(h)^\dagger  \bar{\mathcal{N}}(h)^\dagger \mathcal{E}(p)^\dagger
\mathcal{E}(p) \mathcal{N}(h)  \bar{\mathcal{N}}(h)  \propto  \mathcal{E}(p') \mathcal{N}(h') \bar{\mathcal{N}}(h')$, where
\begin{equation}
	\begin{aligned}
		\mathcal{E}(p') \mathcal{N}(h') \bar{\mathcal{N}}(h') 
		& = \prod_e [(1-p')+p' \sigma_e \bar{\sigma}_e] (1+h' \sigma_e) (1+ h' \bar{\sigma}_e)
		\\ 
		& \propto  \prod_e \Big[ (1+h' \sigma_e) (1+ h' \bar{\sigma}_e) + \frac{p'}{1-p'} (\sigma_e + h') (\bar{\sigma}_e + h') \Big] \\
		& = \prod_e \Big[ (1+h' \sigma_e) (1+ h' \bar{\sigma}_e) + \frac{p'}{1-p'} h'^2(1+ \frac{1}{h'} \sigma_e ) (1  + \frac{1}{h'} \bar{\sigma}_e) \Big] \\ 
	\end{aligned}
\end{equation}
Here $f(p) = 2p(1-p)h'^2/(1-2p+2p^2)$ and $\{S \}$ denotes the set of all distinct collection of edges.
To derive Eq.\eqref{eq:nonfixedZ}, we first note that $\langle \Psi_0, \bar{\Psi}_0 |\prod_e (1+h' \sigma_e)(1+h' \sigma_e) |\Psi_0, \bar{\Psi}_0 \rangle = \sum_{z_\mathbf{v}, t_\mathbf{v}, \bar{z}_\mathbf{v}, \bar{t}_\mathbf{v}} \prod_e \omega_e (z_{\mathbf{v}}, t_{\mathbf{v}}; h)  \bar{\omega}_e (\bar{z}_{\mathbf{v}}, \bar{t}_{\mathbf{v}}; h) $, where $\omega_e(z_{\mathbf{v}}, t_{\mathbf{v}}; h) = \left[\sqrt{2} + h + h (z_v z_{v'} + t_v t_{v'}) + (\sqrt{2} - h) z_v z_{v'} t_v t_{v'} \right]$.
It follows that 
\begin{equation}
	\begin{aligned}
		\langle \rho | \rho \rangle & \propto \langle \Psi_0, \bar{\Psi}_0 | \mathcal{E}(p') \mathcal{N}(h') \bar{\mathcal{N}}(h')  | \Psi_0, \bar{\Psi}_0 \rangle  \\
		& = \sum_{z_\mathbf{v}, t_\mathbf{v}, \bar{z}_\mathbf{v}, \bar{t}_\mathbf{v}} \prod_e \left[\omega_e(z_{\mathbf{v}}, t_{\mathbf{v}}; h) \omega_e(\bar{z}_{\mathbf{v}}, \bar{t}_{\mathbf{v}}; h)  + f(p)\, \omega_e(z_{\mathbf{v}}, t_{\mathbf{v}}; h) \omega_e(\bar{z}_{\mathbf{v}}, \bar{t}_{\mathbf{v}}; h) \right].
	\end{aligned}
\end{equation}

\section{Statistical mechanical mapping for non-fixed-point toric code under phase-flip channel}
\label{sec:chamon}
In this Appendix, we are interested in subjecting non-fixed point toric code wave functions to the phase-flip channel $\mathcal{E}[\cdot] = \prod_e \mathcal{E}_e [\cdot]$, where $\mathcal{E}_e [\cdot] = (1-p)\rho  + p Z_e \rho Z_e$.
Specifically, we choose the initial pure state as the toric code wavefunction evolved by a single-site non-unitary operator with a tuning parameter $h \in [0,1]$:
\begin{equation}
	\label{Eq:chamon_state}
	|\Psi(h)\rangle = \prod_e (1+h Z_e) |\Psi_0\rangle \propto e^{K \sum_e Z_e}|\Psi_0\rangle,
\end{equation}
where $K = \log(1+h) - \log(1-h)$.
We note that Ref.\cite{castelnovo2008quantum} has shown that $|\Psi(h)\rangle$ is a topologically ordered state when  $h <h_c = 1+\sqrt{2} + \sqrt{2(1+\sqrt{2})} \approx 0.217$ while is a short-range-entangled state when $h>h_c$. 
On the other hand, Ref.\cite{fan2023diagnostics} has shown that when $h = 0$, the $n$-th moment of the decohered density matrix $\rho = \mathcal{E}[|\Psi(0)\rangle \langle \Psi(0)|]$ can be mapped to $(n-1)$-th flavored Ising model.
In this Appendix, we will show that for any $p$ and $h$, the $n$-th moment of the density matrix can be mapped to the $n$-flavored Ising model with the $s$-th flavored spin coupled to the $(s+1)$-th (mod $n$) flavored spin through four-spin interactions.
We will also comment about what we expect when taking the limit $n \rightarrow 1$.

To see how the initial pure state density matrix $\rho_h = |\Psi(h)\rangle \langle \Psi(h)|$ evolves under the phase-flip channel, it is more convenient to express $|\Psi(h)\rangle$ in terms of the loop configurations using the fact that $|\Psi_0\rangle \propto \sum_g g|0\rangle \propto \sum_g |g\rangle$ with $g$ denotes the closed-loop Pauli-X operators.
A repeated use of $e^{K Z_e} X_e |0\rangle = X_e e^{-K Z_e} |0\rangle = e^{-K} X_e |0\rangle$ shows that
$|\Psi(h)\rangle= \sum_g e^{-K  |g|} |g\rangle/\sqrt{\mathcal{Z}}$ where $\mathcal{Z} = \sum_g e^{-2K|g|}$ and $|g|$ denotes the total length of the given loops configuration.
Therefore, the initial pure state density matrix takes the form
\begin{equation}
	\rho =\frac{1}{\mathcal{Z}} \sum_{g,\bar{g}} e^{-K (|g| + |\bar{g}|)}|g\rangle \langle \bar{g}|.
\end{equation}
The decohered density matrix after subjecting to channel $\mathcal{E}_e [\cdot] = (1-p)\rho  + p Z_e \rho Z_e$ can then be computed using the similar method as $|\Psi(h)\rangle$, and the result is 
\begin{equation}
	\label{Eq:chamon_dm}
	\rho =\frac{1}{\mathcal{Z} N_G} \sum_{g, \bar{g}} e^{-K (|g| + |\bar{g}|) - K_4 | g \bar{g} |} |g\rangle \langle \bar{g}|,
\end{equation}
where $K_4 = -\log(1-2p)$ and $N_G$ denotes the number of all possible loops configurations. One can now write the $n$-th moment density matrix as
\begin{equation}
	\label{Eq:chamon_rhon}
	\tr(\rho^n) = \frac{1}{(\mathcal{Z}N_G)^n} \sum_{ \{g^{(s)}\} } e^{-2K\sum_{s = 1}^n |g^{(s)}| -K_4 \sum_{s = 1}^n|g^{(s)} g^{((s+1)_n)}|},
\end{equation}
where $(s+1)_n$ denotes the sum modulo $n$.
Eq.\eqref{Eq:chamon_rhon} can be regarded as  $n$ different flavors of loop models, and the given flavor $s$ coupled with the flavor $(s+1)_n$ through the relative loop configurations.
One can also map it to the spin model by identifying the flavor $s$ loop as the domain walls of the flavor $s$ Ising spins, and Eq.\eqref{Eq:chamon_rhon} can be written as $\tr(\rho^n)  =  \sum_{ \{\sigma^{(s)}_\mathbf{v} \} } e^{-H_n}/\mathcal{Z}^n$,
where
\begin{equation}
	\label{Eq:spin_chamon_n}
	-H_n  =  \sum_{ \langle v,v'\rangle }\sum_{s = 1}^n  \big( 2 K  \sigma^{(s)}_{v} \sigma^{(s)}_{v'}+K_4    \sigma^{(s)}_{v} \sigma^{(s)}_{v'}   \sigma^{(s+1)_n}_{v} \sigma^{(s+1)_n}_{v'} \big).
\end{equation}
Therefore, we have mapped the $n$-th moment density matrix  as the n-flavored Ising model with flavor $s$ spin coupled with the flavor $(s-1)_n$ and $(s+1)_n$ through the four-spin interactions.
We note that for $n = 2$ and $3$, $H_n$ is equivalent to the 2-flavored and 3-flavored \cite{grest1981n} Ashkin-Teller (AT) model.
However, we are not aware of any previous work that has studied $H_n$ for  $n \geq 4$.

We now focus on the case where $n = 2$, whose error threshold is expected to be the closest to the $n \rightarrow 1$ limit compared to other integer R\'enyi indices.
The exact mapping to the isotropic Ashkin-Teller model allows us to determine the phase boundary and the universality of the transition as shown in 
Fig.\ref{Fig:chamon_phase}(a). 
One can similarly use the anyon condensation and confinement paremeters mentioned in the main text to identify three phases: the mixed-state topologically ordered phase, the classical topologically ordered phase, and the trivial phase.
We note that the topologically ordered phase is mapped to the paramagnetic phase of the AT model, which is different from the mapping for the fixed-point toric code subjected to coherent channel.
Another interesting feature of the phase diagram is that when $h$ is large but below the critical point of the pure-state transition, one can increase $p$ and drive the system from topological to the trivial phase through the BKT transition.
However, we believe this phenomena is due to the replica index and will disappear when $n \rightarrow 1$, since the Ising$^2$ transtion along the $p = 0$ line (i.e., the pure-state transition line) becomes the usual Ising transtion when $n \rightarrow 1$.

Finally, we breifly comment on the expected phase diagram in the $n \rightarrow 1$ limit.
In this limit, both the $p = 0$ and $p = 0.5$ lines can be exactly mapped to the 2d classical Ising model, and crucially, they have the same critical field strength $h_c = 1+\sqrt{2} + \sqrt{2(1+\sqrt{2})} \approx 0.217$.
Therefore, we expect a vertical boundary with 2d Ising transition at $h = h_c$.
On the other hand, the $h = 0$ limit can be exactly mapped to the random-bond (RB) Ising model along the Nishimori line, where the transtion occurs at $[ = p_c \approx 0.109$.
Therefore, one expects a smooth line with RB Ising transition connecting $(h,p) \approx (0,0.109)$ and $(h,p) \approx (0.217,0)$.
We summaize the expected $n \rightarrow 1$ phase diagram in Fig.\ref{Fig:chamon_phase} in the main text.

\clearpage

\end{document}